\documentclass[preprint, twocolumn]{aastex631}
\usepackage{natbib}[1999/02/25]
\usepackage{booktabs}
\usepackage{amsmath}
\usepackage{CJKutf8}

\usepackage{psfrag,color}
\usepackage{multirow}
\usepackage{subfigure}

\newcount\ppnum
\newcommand\ppnumber[1]{%
        \ppnum=#1\relax
        \ifnum\ppnum<0
                $-$%
                \ppnum=-\ppnum
        \fi
        \let\pptemp\empty
        \loop\ifnum\ppnum>999
                \count255=\ppnum
                \divide\ppnum by1000
                \count255=\numexpr \count255 - 1000*\ppnum \relax
                \edef\pptemp{,\ifnum\count255<100 0\ifnum\count255<10 0\fi\fi
                             \the\count255 \pptemp}%
        \repeat
        \the\ppnum
        \pptemp
}
\shorttitle{CN Lyn}
\shortauthors{Y\"{u}cel et al.}
\graphicspath{{./}{Figures/}}

\begin{document}

\title{Unveiling the Origins and Dynamics of the Hierarchical Triple Star System CN Lyn}	
\correspondingauthor{G\"{o}khan Y\"{u}cel}
\email{gokhannyucel@gmail.com}

\author[0000-0002-9846-3788]{G\"{o}khan Y\"{u}cel}
\altaffiliation{T\"{U}B{\.{I}}TAK-2218 Fellow}
\affiliation{Department of Astronomy~\& Space Sciences, Istanbul University, 34119, Istanbul, T\"{u}rkiye}
	
\author[0000-0002-3125-9010]{Volkan Bak{\i}\c{s}}
\affiliation{Department of Space Sciences~\& Technologies, Akdeniz University, 07058, Antalya, T\"{u}rkiye}

\author[0000-0003-2575-9892]{Remziye Canbay}
\affiliation{Institute of Graduate Studies in Science, Istanbul University, 34116, Istanbul, T\"{u}rkiye}
	
\author[0000-0001-9809-7493]{Neslihan Alan}
\affiliation{Department of History of Science, Fatih Sultan Mehmet Vakif University, 34664, Istanbul, T\"{u}rkiye}

\author[0000-0001-9445-4588]{Timothy Banks}
\affiliation{Nielsen, 675 6th Ave, New York 10011, USA}
\affiliation{Harper College, 1200 W Algonquin Rd, Illinois 60067, USA}

\author[0000-0003-3510-1509]{Sel\c{c}uk Bilir}
\affiliation{Department of Astronomy~\& Space Sciences, Istanbul University, 34119, Istanbul, T\"{u}rkiye}
		
	
	
\begin{abstract}
\noindent In this study we present a detailed analysis of CN Lyn, an overlooked triple star system, by combining spectroscopic data from the literature, photometric \textit{TESS} data, and kinematic techniques. We updated the fundamental parameters of the known eclipsing components in the system with high precision. The chemical composition of both eclipsing components (Aab) and the third component (B) in the system were calculated with great accuracy. According to our analysis the mass, radius, and metallicity of the eclipsing components are $1.166_{-0.012}^{+0.013}\,M_\odot$, $1.786_{-0.014}^{+0.013}\,R_\odot$, and $-0.78_{-0.02}^{+0.02}$ dex for Aa and $1.143_{-0.012}^{+0.013}\,M_\odot$, $1.651_{-0.013}^{+0.014}\,R_\odot$, and  $-0.55_{-0.02}^{+0.03}$ dex for Ab. The pair's age is $3.89_{-0.10}^{+0.10}$ Gyr. The mass, radius, metallicity, and age for B are $0.85_{-0.23}^{+0.23}\,M_\odot$, $1.436_{-0.023}^{+0.026}\,R_\odot$, $-1.83_{-0.11}^{+0.09}$ dex, and $12.5_{-2.5}^{+2.5}$ Gyr, respectively.  It is also found that the triple system (AabB) satisfies the stability criteria for the hierarchical triple system. Kinematic and Galactic orbital parameters of CN Lyn were obtained from the astrometric and spectroscopic data of the system. Dynamical orbital analyses, taking into account the ages of the component stars in the central binary system (A) show that the CN Lyn originated at the metal-poor edge of the Galactic disk. The third component of the system was found to be a member of the halo population in terms of age, $\alpha$ elements and metal abundance. Given the different chemical abundances and age of B compared to A, this suggests that the third component was captured by the central system in a region with weak gravitational interactions far beyond the Galactic disc.


\end{abstract}
	
	

 
\section{Introduction} \label{sec:intro}

Triple star systems hold significant astrophysical importance, offering profound insights into stellar evolution, dynamical interactions, and mass transfer processes. Due to their complexity compared to single or binary star systems, triple systems facilitate a more comprehensive understanding of stellar processes from formation to final stages. Events such as mass transfer and mergers within such systems can trigger high-energy astrophysical phenomena, including supernova explosions and the formation of compact objects like neutron stars and black holes \citep{Toonen2020}. The dynamical stability of triple systems, characterized by interactions between inner and outer orbits, contributes critically to the study of stellar dynamics \citep{Eggleton2001}. Additionally, triple systems serve as important sources of gravitational waves, offering valuable insights into this area of astrophysical research \citep{Antonini2012}.

The distribution of triple star systems across different spectral classes sheds light on their formation and evolutionary processes. Spectral classification, which organizes stars based on temperature, luminosity, and spectral lines, reveals distinct patterns in the occurrence and characteristics of triple systems. Observational studies have identified triple star systems across all spectral classes, from hot O-type stars to cooler M-type stars, though their distribution is uneven. To illustrate: massive O- and B-type stars are more frequently found in higher-order multiple systems, including triples, likely due to their formation in dense stellar environments where dynamical interactions are more prevalent \citep{Sana2014, Moe2017}. In contrast to the high-mass stars, middle/low-mass stars (such as those in the F, G, K, and M spectral classes) less commonly form triple systems. When they do, these systems typically exhibit wider separations and lower eccentricities than the early type systems \citep{Tokovinin2008}. 

The frequency of triple star systems also varies with metallicity and age, influencing their spectral distribution. Metal-rich environments tend to favor the formation of tighter, more stable multiple systems, while metal-poor environments are more likely to produce wider, less bound systems \citep{Raghavan2010}. Additionally, older star populations display a different distribution compared to those found in, say, younger clusters, reflecting the dynamical evolution and potential disruption of triple systems over time. The spectral class distribution of triple star systems underscores the complex interplay between stellar mass, formation environment, and dynamical evolution, providing a thorough understanding of their occurrence and characteristics across the Hertzsprung-Russell (HR) diagram. Studying these systems enables researchers to gain deeper insights into the mechanisms governing star formation, the stability of multiple star systems, and the ultimate outcomes of their evolutionary paths. Therefore, triple-star systems are essential for advancing our knowledge in stellar astrophysics.

Orbital parameter studies on tertiary components in triple systems have interesting findings. \cite{Kyle2014} has analyzed the \textit{Kepler} Eclipsing Binary Catalog\footnote{\url{https://keplerebs.villanova.edu/}} for third bodies in eclipsing binaries. Even though they have been limited by the operational time-span of \textit{Kepler} ($\sim$\ppnumber{1400} days) they found that among \ppnumber{1279} close eclipsing binaries there are 236 tertiary components, which aligns with the results of \cite{Tokovinin2006} that $\sim15\%$ of binaries have a third component. \cite{Borkovits2016} included systems with longer periods, which were excluded by \cite{Kyle2014}, and extended the number of systems from the \textit{Kepler} Eclipsing Binary Catalog. According to their analyses, there are 104 triple candidates with the outer orbital period being less than \ppnumber{1000} days and 105 triple candidates with periods less than \ppnumber{1500} days. The distribution of triple candidates with 200 $\leq P$ (day) $\leq$ \ppnumber{1600} is almost flat but the number of the system with longer periods, which contains a tertiary component, rapidly declines.

CN Lyn (GSC 02973-00253, SAO 60517, Hip 39250, $l=181^{\rm o}\!\!.759605,~~b=29^{\rm o}\!\!.646774$), the subject of this study, was detected by the HIgh Precision PARallax COllecting Satellite \citep[\textit{Hipparcos,}][]{Perryman} as a detached eclipsing binary system with an orbital period of 1.9554 days. This binary system is notable for the inconsistency between its luminosity class as determined from photometry and the classification derived from \textit{Hipparcos} parallax data \citep{Grenier1999}. This discrepancy was taken as an indicator of CN Lyn being a multiple system. The first ground-based observation of CN Lyn was made by \cite{Marrese2004} via obtaining high-resolution spectra for the purpose of evaluating \textit{Gaia} performance \citep[c.f.,][]{Munari2001, Zwitter2003}. Combining spectroscopic parameters and photometric solutions, \citet{Marrese2004} determined the physical parameters of the CN Lyn with high precision, confirming that it is a triple system. The effective temperatures were found to be $T_{\rm eff,1}=\ppnumber{6500}\pm250$ K and $T_{\rm eff,2}=\ppnumber{6455}\pm260$ K. The masses of the inner pair of stars (Aab) were derived as $M_{\rm 1, 2}=1.04\pm0.02 M_{\odot}$, and the radii as $R_{\rm 1}=1.80\pm0.21 R_{\odot}$, $R_{\rm 2}=1.84\pm0.24 R_{\odot}$. Additionally, the mean radial velocity of the third body (B) was determined to be $V_{\rm R}=-13\pm1$ km s$^{-1}$, which is in agreement with the systemic velocity of the CN Lyn as $V_{\gamma}=-15.6\pm0.5$ km s$^{-1}$. Light from the third star in the system can be clearly seen on the spectra they obtained \citep[see Fig. 1 in][]{Marrese2004}. They also stated that star B has a light contribution of 29 $\pm$ 6\% in the \textit{Hipparcos}-$H_{\rm p}$ band, and that the third companion probably is identical to other components of CN Lyn and might have evolved from the main-sequence. CN Lyn, which is included in a comprehensive catalogue of detached double-lined eclipsing binaries compiled from the literature \citep{Eker2014}, was used by \citet{Eker2015, Eker2018, Eker2024} to derive mass-luminosity relations for main-sequence stars. Recently, new eclipse timing data were obtained by \cite{Liao2021} to determine the orbit of the third companion via orbital period analysis. As a result, the period of the third companion was estimated as being $15.80 \pm 0.40$ years, with a projected semi-major axis of $0.0074 \pm 0.0008$ day, and an eccentricity of $0.38 \pm 0.18$. Assuming the third companion has a similar mass as the components of CN Lyn, they calculated the orbital inclination of the third companion as $25^\circ$. \cite{Liao2021} also obtained a low-resolution spectrum of CN Lyn and based on that data, they determined the atmospheric model parameters of the second component of the system as $T_{\rm eff,2}= \ppnumber{6337}\pm 37$ K, $\log g = 4.27\pm 0.08$ cgs and ${\rm [Fe/H]}_2=-0.67 \pm 0.06$ dex.

In this study, we combined spectra of CN Lyn such as from \cite{Marrese2004}, high-resolution ELODIE spectra, and photometric data from the Transiting Exoplanet Survey Satellite \citep[\textit{\!TESS},][]{{Ricker2015}} to obtain the fundamental parameters of each component of CN Lyn with greater accuracy than previously achieved. We disentangled the spectra of each component, performed a detailed spectral analysis, and accurately determined the temperatures and chemical abundances of up to 20 elements for the eclipsing components of CN Lyn for the first time in the literature. Additionally, we investigated the system's evolutionary scenarios, determining its initial orbital periods, kinematics and Galactic orbital parameters, which provide insights into the birthplace of CN Lyn.

This paper is structured as follows. In Section~\ref{sec:observation} we present the properties of the observational data used. Section~\ref{sec:analysis} describes the newly acquired orbital elements of CN Lyn and the calculated fundamental parameters of the triple system. We present a detailed evolutionary analysis of the CN Lyn in Section~\ref{sec:evolution}, while in Section~\ref{sec:kinematics} we cover the space velocity components and the Galactic orbital parameters of CN Lyn. Finally, we discuss our results in Section~\ref{sec:discussion}.


\section{Observations} \label{sec:observation}


\subsection{Spectroscopic Data}

Our spectroscopic data were taken from two different sources:
\begin{itemize}
    \item{One set of spectra was obtained by \cite{Marrese2004} with an Echelle CCD spectrograph on the 1.82-m telescope operated by the Osservatorio Astronomico di Padova atop Mt.\ Ekar (Asiago). A total of 29 spectra were obtained between 1999 and 2003, each covering the range \ppnumber{4550}~$<\lambda ({\rm \text{\AA}})<$~\ppnumber{9000}. The dispersion was 0.25~$\mathring{\text{A}}$ per pixel which, with a slit width of 2.0 arcsec, leads to a resolution of 0.42~$\mathring{\text{A}}$ or equivalently to a resolving power of $R \sim$~\ppnumber{17000}.}
    \item{The second set of 12 spectra were taken from ELODIE archive\footnote{\url{http://atlas.obs-hp.fr/elodie/}} \citep{elodie}. The ELODIE spectrograph \citep{Baranne}, attached to a 1.93-m reflector telescope at the Observatoire de Haute-Provence. These spectra have the range \ppnumber{3850}~$ < \lambda ({\rm \text{\AA}} ) < $~\ppnumber{6800}, in 67 orders, with a resolving power of $R \sim$ \ppnumber{42000}, and were obtained between 2001 and 2005.}
\end{itemize}  


\subsection{Photometric Data}

\begin{figure*}
    \centering
        \begin{subfigure}
        \centering
        \includegraphics[width=0.9\textwidth]{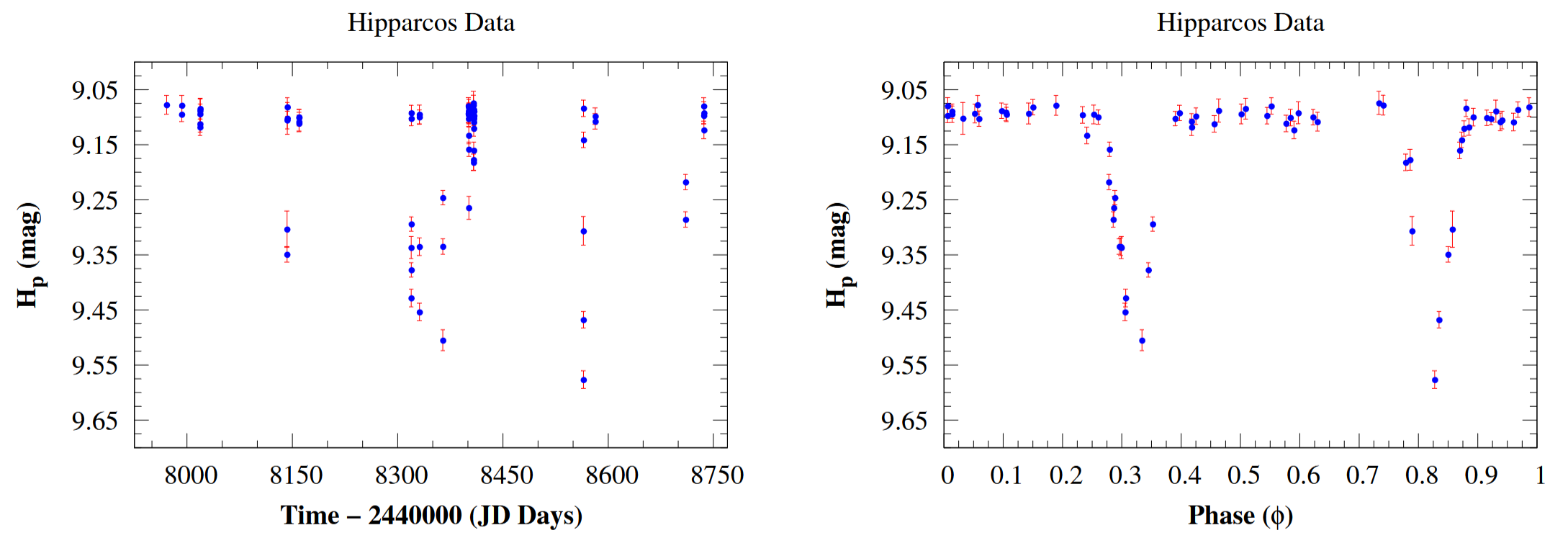}
    \end{subfigure}
    \begin{subfigure}
        \centering
        \includegraphics[width=0.9\textwidth]{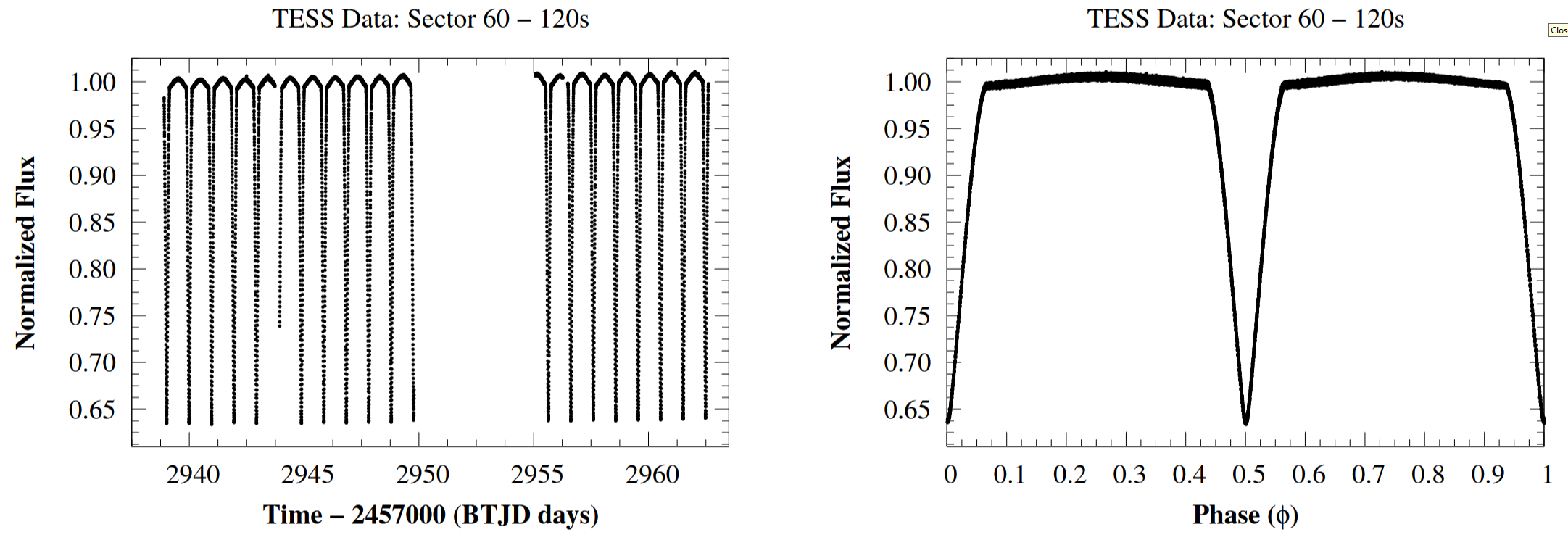}
    \end{subfigure}

    \caption{Upper panel: \textit{Hipparcos} photometric data for CN Lyn. The left subfigure plots $H_{\rm P}$ photometry \citep{van_Leeuwen_1997} against Julian Date (JD). The right subfigure plots the same data phase folded using the ephemeris of \cite{Kreiner_2004}. Lower panel: \textit{TESS} photometric data for CN Lyn. The left subfigure plots flux against Barycentric \textit{TESS} Julian Date (BTJD). The right subfigure shows the same data folded using the ephemeris of \cite{Kreiner_2004}.}
    \label{tess_obs}
\end{figure*}

The photometric data analysed in the current study come from three sources:
\begin{itemize}

    \item{\textit{Hipparcos} observed CN Lyn on dates between February 1990 and May 1992, resulting in $71$ data points. \textit{Hipparcos} epoch photometric data were obtained via CDS \citep{Perryman, Wenger2000}. The \textit{Hipparcos} data used in this study are plotted in the top panel of Figure~\ref{tess_obs}.} 
  
    \item{\textit{TESS:} the main object of the \textit{TESS} space telescope mission \citep{Ricker2015} is to find exoplanets orbiting bright stars, however, its photometric time-series database is also a valuable resource for the study of variable stars such as eclipsing binaries \citep{Southworth_2021, PrsaTESS}.  \textit{TESS} has observed CN Lyn in only one sector, Sector-60\footnote{At the time of paper submission.}, which dates between 23 December 2022 and 18 January 2023, with exposure times of 120 seconds. We used \texttt{lightkurve} v2.4 \citep{lk} code to obtain \textit{TESS} data with the quality flag of ``hard''. We chose simple aperture photometry data from the ``TESS-SPOC'' data reduction. The \textit{TESS} data used in this study are plotted as Figure~\ref{tess_obs}.} 
    
    \item{An additional 13 eclipse timing data points were taken from \cite{Liao2021} based on CCD observations taken over October 2016 to April 2021 by those authors, as well as a further four they reported from Super Wide Angle Search for Planets \citep[SuperWASP;][]{Pollacco2006} and the Kamogata/Kiso/Kyoto Wide-field Survey\footnote{\url{http://kws.cetus-net.org/~maehara/VSdata.py}.}} 
    
\end{itemize}

\begin{table*}[t]
    \caption{Observed times of minima (ToM), epoch number and O--C values calculated using the ephemeris given by \cite{Kreiner_2004}.}
    \centering
    \scriptsize
  	\renewcommand{\arraystretch}{0.75}
    {\begin{tabular}{lccccccccc}
    \hline
       ToM       & Method & Epoch No & O-C          & Reference & ToM       & Method & Epoch No & O-C       & Reference  \\
       (HJD)      &       &          & (days)       &           & (HJD)     &        &          & (days)    &            \\    
    \hline
    2448500.25400 &  Pe   & 0       & 0             & 1     & 2458219.13423 &  CCD  & 4970      & 0.0005        & 14 \\
    2451450.14700 &  CCD  & 1508.5  & 0.0077        & 2     & 2458427.39507 &  CCD  & 5076.5    &  $-$0.0004    & 14 \\
    2454116.47714 &  CCD  & 2872    & 0.0013        & 3     & 2458569.16626 &  CCD  & 5149      &  $-$0.0036    & 14 \\
    2454518.33330 &  Pe   & 3077.5  & 0.0004        & 4     & 2458570.14613 &  CCD  & 5149.5    &  $-$0.0015    & 14 \\
    2454556.46490 &  CCD  & 3097    &$-$0.0005      & 5     & 2458823.38035 &  CCD  & 5279      &  $-$0.0057    & 14 \\
    2454556.46560 &  CCD  & 3097    & 0.0002        & 5     & 2458827.28935 &  CCD  & 5281      &  $-$0.0077    & 14 \\
    2454556.46730 &  CCD  & 3097    & 0.002         & 5     & 2459317.14394 &  CCD  & 5531.5    &  $-$0.0081    & 14 \\
    2455871.53590 &  CCD  & 3769.5  &  $-$0.0093    & 6     & 2459318.12480 &  CCD  & 5532      &  $-$0.0050    & 14 \\
    2456325.21740 &  CCD  & 4001.5  &  $-$0.0058    & 7     & 2459939.00106 &  CCD  & 5849.5    & 0.0001        & 16 \\
    2456400.50530 &  CCD  & 4040    &  $-$0.0051    & 7     & 2459939.97891 &  CCD  & 5850      & 0.0002        & 16 \\
    2456404.41650 &  CCD  & 4042    &  $-$0.0049    & 8     & 2459940.95656 &  CCD  & 5850.5    & 0.0001        & 16 \\
    2456646.89700 &  CCD  & 4166    &  $-$0.0075    & 7     & 2459941.93445 &  CCD  & 5851      & 0.0002        & 16 \\
    2456670.36850 &  CCD  & 4178    &  $-$0.0021    & 9     & 2459942.91212 &  CCD  & 5851.5    & 0.0002        & 16 \\
    2456709.47720 &  CCD  & 4198    &  $-$0.0036    & 10    & 2449944.86757 &  CCD  & 5852.5    & 0.0001        & 16 \\
    2456711.43100 &  CCD  & 4199    &  $-$0.0053    & 10    & 2459945.84546 &  CCD  & 5853      & 0.0002        & 16 \\
    2456712.41020 &  CCD  & 4199.5  &  $-$0.0038    & 10    & 2459946.82309 &  CCD  & 5853.5    & 0.0001        & 16 \\
    2456713.38790 &  CCD  & 4200    &  $-$0.0039    & 9     & 2459947.80098 &  CCD  & 5854      & 0.0002        & 16 \\
    2456714.36480 &  CCD  & 4200.5  &  $-$0.0048    & 10    & 2459948.77861 &  CCD  & 5854.5    & 0.0002        & 16 \\
    2457056.58120 &  CCD  & 4375.5  &  $-$0.0024    & 11    & 2459949.75642 &  CCD  & 5855      & 0.0002        & 16 \\
    2457057.56020 &  CCD  & 4376    &  $-$0.0012    & 11    & 2459955.62306 &  CCD  & 5858      & 0.0001        & 16 \\
    2457409.55250 &  CCD  & 4376    &  $-$0.0012    & 12    & 2459956.60070 &  CCD  & 5858.5    & 0.0001        & 16 \\
    2457815.32460 &  CCD  & 4376    &  $-$0.0012    & 13    & 2459957.57859 &  CCD  & 5859      & 0.0001        & 16 \\
    2457681.36994 &  CCD  & 4695    & 0.0012        & 14    & 2459958.55619 &  CCD  & 5859.5    & 0.0001        & 16 \\
    2457820.21171 &  CCD  & 4766    & 0.0018        & 14    & 2459959.53408 &  CCD  & 5860      & 0.0001        & 16 \\
    2458123.31537 &  CCD  & 4921    & 0.0016        & 14    & 2459960.51173 &  CCD  & 5860.5    & 0.0001        & 16 \\
    2458125.27054 &  CCD  & 4922    & 0.0012        & 14    & 2459961.48962 &  CCD  & 5861      & 0.0001        & 16 \\
    2458155.58140 &  CCD  & 4937.5  & 0.0017        & 15    & 2459962.46731 &  CCD  & 5861.5    & 0.0001        & 16 \\
    2458178.07376 &  CCD  & 4949    & 0.0057        & 14   \\
    \hline
    \end{tabular}}
    \\
    \footnotesize {\raggedright References: (1)  \cite{ESA1997}, (2) Anton Paschke, (3) \cite{Pollacco2006}, (4) \cite{Yilmaz2009}, (5) \cite{Brat2008}, (6) \cite{Honkova2013}, (7) \cite{KWS}, (8) \cite{Honkova2014}, (9) \cite{Hubscher2014}, (10) \cite{Hubscher2015}, (11) \cite{Hubscher2015-2}, (12) \cite{Hubscher2017}, (13) \cite{Pagel2018}, (14) \cite{Liao2021}, (15) Agerer Franz, (16) This study \par}
    \label{tab:minimas}
\end{table*}

\section{Data Analyses} \label{sec:analysis}


\subsection{Orbital Period Analysis and the Third Body Orbit}

A total of 55 times of minima are available in the literature, distributed between 1991 and 2023, with the data being well distributed after 2007. These timings are given in Table~\ref{tab:minimas}, which starts with those based on \textit{Hipparcos} \citep{ESA1997} observations and ends with timings from \textit{TESS} \citep{Ricker2015} observations. 

\begin{figure*}
  \centering
	\includegraphics[scale=0.35]{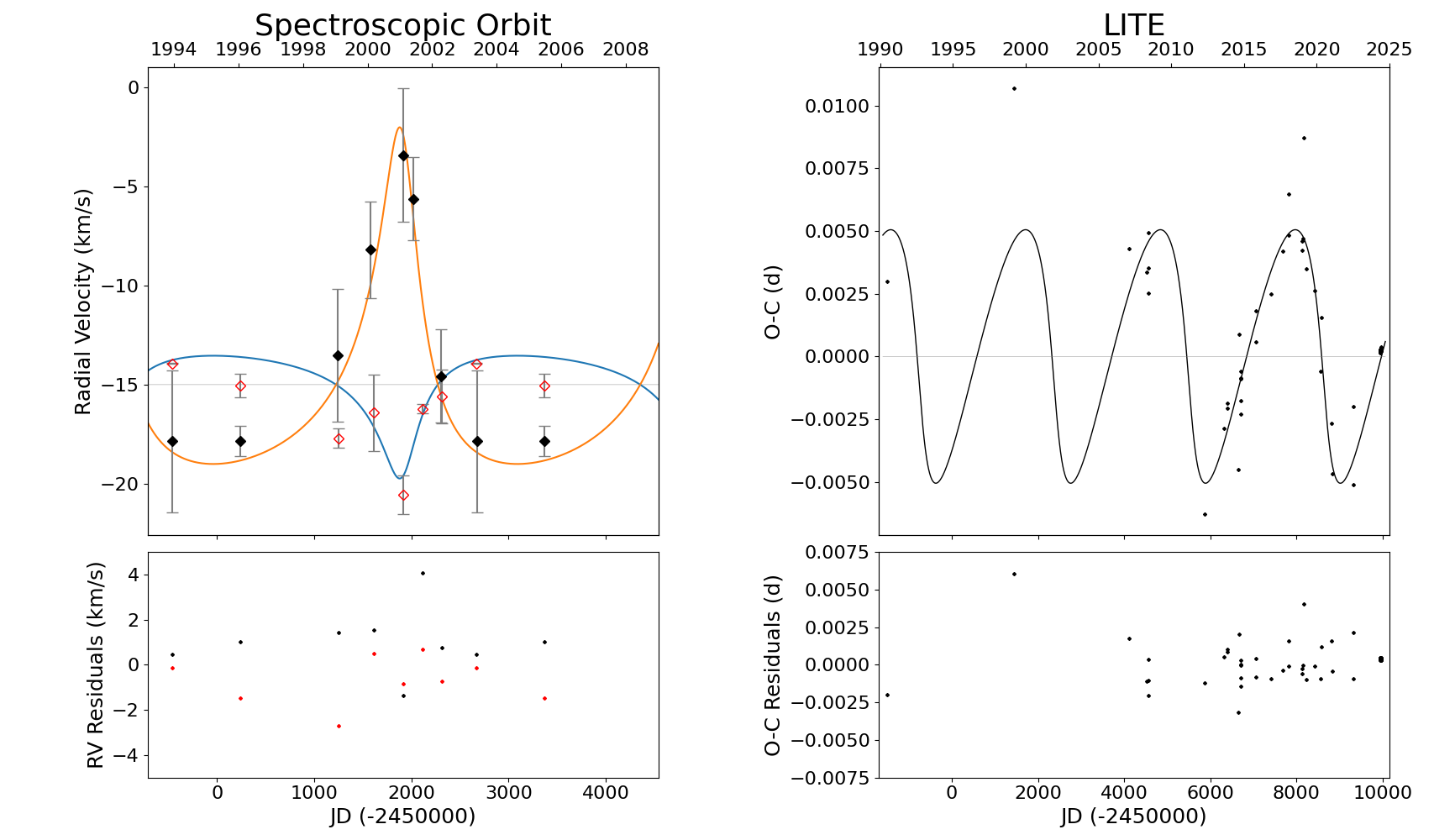} \\
    \vspace{5mm}
	\includegraphics[scale=0.25]{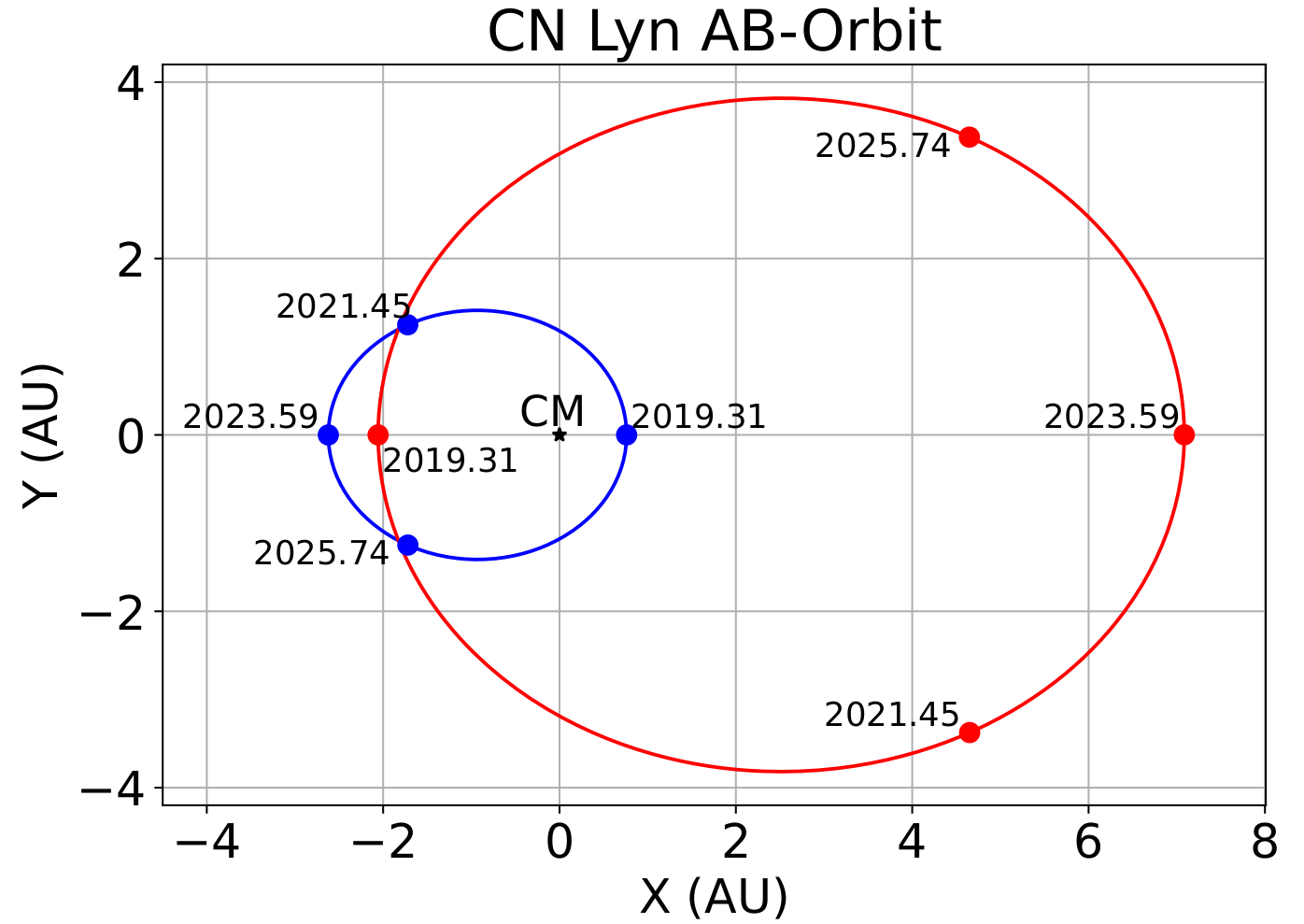}
        \caption{\textit{Upper panel}: RV and O-C variation due to orbital motion with the third body. \textit{Lower panel:} The components' orbit and location at specific dates. Blue and red continuous lines show the orbits of Aab and B components, respectively, and CM represents the center of mass of the CN Lyn.
        \label{cnlyn_rvorbit}
    }
\end{figure*}

Figure~\ref{cnlyn_rvorbit} plots the  O--C data using the linear ephemeris from the \citet{ESA1997}. This shows a clear quasi-sinusoidal variation suggestive of a light-time effect (LITE, \citealt{Irwin_1952, Irwin1959}) associated with a third-body. In order to analyze these O--C data, we applied the LITE formula given by \cite{Irwin1959}:

\begin{eqnarray}
    T(E) = A + BE + C E^2 +  \frac{a_{12}\sin{i_3}} {c} \times ~~~~~~~~~~~~~~~~~~~~~~~~~~\\ 
    \left( \frac{1-e_3^2}{1+e_3\cos{\nu_3}}\sin(\nu_{3} + \omega_{3}) + e_{3}\cos{\omega_3} \right) \nonumber
    \label{omega_obs}    
\end{eqnarray}   

\noindent
where $c$ is the speed of light, and the other symbols have their usual meanings as given by \cite{Irwin1959}. A weighted least-squares solution for $T_0$, $P_{\rm orb}$, $Q$, $a_{\rm 12}$, $i_3$, $e_3$, $\omega_3$, $T_3$ and $P_{\rm 12}$ is presented in Table~\ref{tab:big_orbit}. The upper panel of Figure~\ref{cnlyn_rvorbit} displays the observational data together with the theoretical best-fit curves, together with the residuals from the fits.

\begin{table}
    \caption{LTE parameters for the model fit to the AB orbit.}
    \centering
    \begin{tabular}{lr}
    \hline
    Parameters                                  & Value\\
    \hline
     $T_0$   (HJD)                              & $48500.251 \pm 0.003$   \\
     $P$ (d)                                    & $1.9555091 \pm 0.0000015$   \\
     $Q$                                        & 0.0   \\
     $a_\mathrm{12}\sin i_\mathrm{12}$ (AU)     & $1.03 \pm 0.14$\\
     $e_\mathrm{3}$                             & $0.55 \pm 0.14$ \\
     $w_\mathrm{3}$ (deg)                       & $196 \pm 8$ \\
     $T_\mathrm{3}$ (HJD)                       & $43017 \pm 410$\\
     $P_\mathrm{3}$ (d)                         & $3130 \pm 78$ \\
     $f(m_\mathrm{3})\left(\frac{(a_{12} \sin{i_{12}})^3}{P_{3}^2}\right)$ ($M_\odot$)     & $0.0149\pm0.0081$\\
    \hline
    \end{tabular}
    \label{tab:big_orbit}
\end{table}

Three components are particularly noticeable in the composite spectrum of CN Lyn. This is especially clear for the Balmer lines (see Figure~A1, which has been placed into the appendix given its size) where the blending of individual lines is maximum. Blending of spectral lines shifts the line cores, resulting in inaccurate radial velocity (RV) measurements. To address this, we developed a program code that simultaneously fits the observed spectrum with the composite of three synthetic spectra that are shifted in the wavelength axis. The atmospheric parameters for the synthetic spectra are selected based on the physical properties of the components as \ppnumber{6500}~K, 4.0 cgs, $-0.5$ dex for temperature, surface gravity, and metallicity, respectively. We used the \texttt{ATLAS9} \citep{Castelli2004} code to produce atmosphere models and \texttt{SPECTRUM} \citep{Gray1994} to produce synthetic spectra.

A numerical optimization technique has been employed to estimate uncertainties in the shift parameters. The method involves minimizing a chi-square function that quantifies the difference between observed and composite spectra, where the composite spectrum is generated by linearly combining template spectra shifted by varying amounts. Specifically, we utilized the Nelder-Mead algorithm \citep{Nelder, press2007} implemented in the \texttt{scipy.optimize} module of Python, which iteratively adjusts shift parameters to minimize the discrepancy between observed and model spectra. Uncertainties in the optimal shift parameters were estimated by exploring the variation in chi-square values as the optimization process converges to different local minima. This approach provided us a robust estimate of parameter uncertainties around 0.2 km~s$^{-1}$. It should be noted that the uncertainties determined from the fitting process primarily account for the statistical errors inherent in the model fitting procedure. However, the data itself is subject to additional sources of uncertainty due to wavelength calibration errors. These calibration errors arise from imperfections and variances in the measurement instruments and processes, contributing to a systematic uncertainty in the final results. Therefore, the total uncertainty in the analysis is a combination of the fitting uncertainties and the calibration errors of the data, and both need to be considered to provide a comprehensive assessment of the reliability of the findings. The best-fitting matches for each observed spectrum are shown in Figure~A1, and the measured RVs are provided in Table~\ref{table_RVs}. The analysis of the RVs of the primary and secondary components yielded the elements of the spectroscopic orbit of the close binary (Aab) given in Table~\ref{tab:spec_orbs_a}, close to the literature values \citep{Marrese2004}. 

Plotting the tertiary component's RVs over time reveals a quasi-sinusoidal variation similar to the O--C data. To determine if the systemic velocity of Aab changes due to the tertiary component orbiting about them, we divided the RV dataset into seven intervals. Each interval was analyzed by converging only the systemic velocity parameter, resulting in a set of seven systemic velocities that vary in accordance with the tertiary component's RVs (see Figure~\ref{cnlyn_rvorbit}) which suggests a possible third-body orbit. Fixing the orbital period, eccentricity, and the longitude of periastron passage to the estimated values from the O--C analysis, we fitted a spectroscopic orbit for a possible Aab + B solution, obtaining a model compatible with the O--C analysis. This analysis yielded $K_\mathrm{Aab}=3.1\pm 0.7$ km s$^{-1}$ and $K_\mathrm{B}=8.5\pm 0.8$ km s$^{-1}$ (mass ratio $q(K_\mathrm{Aab}/K_\mathrm{B})=0.37$, see also Table~\ref{tab:spec_obs_AB} for other orbital parameters). It should be noted here that the projected semi-major axis ($a_{\rm A}$sin $i_{\rm AB}=0.76\pm0.18$) from the spectroscopic orbit of the triple system is $\sim1.5\sigma$ smaller than the LITE orbit obtained from the O--C analysis ($a_{\rm A}$sin $i_{\rm AB}=1.03\pm0.14$) causing a lower third-body mass estimate. In the Discussion section, the mass probability of the third star is discussed with evidence of its metallicity and evolutionary status. The inconsistency in the projected semi-major axis between spectroscopic and LITE orbital solutions may be due to long-term changes in the orbital period \citep[e.g.][]{Liao2021}, which is unclear in the present O--C data coverage. In Figure~\ref{cnlyn_rvorbit}, the orbits in the AB system and the components' location at specific dates are shown.

\begin{table}
    \caption{Spectroscopic parameters of the wide orbit (AB-system).}
    \centering
    \begin{tabular}{lc}
    \hline
    Parameters                                  & Value\\
    \hline
     $P_\mathrm{AB}$ (d)                                    & 3130$\pm$78\\
     $T_0$                                      & 48783$\pm$26\\
     $K_\mathrm{A}$ (km\,s$^{-1}$)              & 3.1$\pm$0.7 \\
     $K_\mathrm{B}$ (km\,s$^{-1}$)              & 8.5$\pm$0.8 \\
     \emph{e}                                   & 0.55$\pm$0.14\\
     \emph{w} ($^\circ$)                        & 196$\pm$8\\
     $V_{\gamma {\rm AB}}$ (km\,s$^{-1}$)       & --14.9$\pm$0.4\\
     $m_\mathrm{A}\sin{^3{i_\mathrm{AB}}}$ ($M_\odot)$      & 0.215$\pm$0.040\\
     $m_\mathrm{B}\sin{^3{i_\mathrm{AB}}}$ ($M_\odot)$      & 0.080$\pm$0.040\\
     \emph{q} ($M_\mathrm{B}/M_\mathrm{A}$)     & 0.37$\pm$0.09\\
     $a_\mathrm{A}\sin{i}_\mathrm{AB}$ (AU)    & 0.755$\pm$0.175\\     
     $a_\mathrm{B}\sin{i}_\mathrm{AB}$ (AU)    & 2.034$\pm$0.192\\
     $f(m_\mathrm{A})\left(\frac{(a_\mathrm{B} \sin{i_\mathrm{AB}})^3}{P_\mathrm{AB}^2}\right)$ ($M_\odot$)    & 0.1146$\pm$0.0434\\     
     $f(m_\mathrm{B})\left(\frac{(a_\mathrm{A} \sin{i_\mathrm{AB}})^3}{P_\mathrm{AB}^2}\right)$ ($M_\odot$)    & 0.0058$\pm$0.0057\\
     
     rms (km s$^{-1}$)                          & 1.26 \\
     \hline
    \end{tabular}
    \label{tab:spec_obs_AB}
\end{table}

Having the spectroscopic orbit of the triple system allowed us to model the RVs of the close binary components, which are influenced by both the third-body orbit (AB) and their close orbit (Aab). The RVs of the Aab system provided in Table \ref{table_RVs} are systemic velocity-corrected. These refined RVs for the close binary prompted us to revisit our spectroscopic orbital parameters for the system. A re-analysis of the close binary's RVs resulted in the final and more reliable orbital parameters, as shown in Table~\ref{tab:spec_orbs_a}.

\begin{table}[!th]
  \centering
  \scriptsize
  	\renewcommand{\arraystretch}{0.75}
  \caption{Systemic velocity corrected RVs of primary and secondary components (Aab). RVs of the tertiary component (B) refer to measurements from the H$_\beta$ line (see Figure A1).}
    \begin{tabular}{rclrrr}
    \hline
    ID    & HJD           & Phase       & Primary       & Secondary     & Tertiary      \\
          &               &             & (km\,s$^{-1}$) & (km\,s$^{-1}$) & (km\,s$^{-1}$) \\
    \hline
1	&	2451225.46270	&	0.60721	&	$-$80.62	&	64.38	    &	$-$20.59	\\
2	&	2451229.40920	&	0.62536	&	$-$79.74	&	79.26	    &	$-$9.73	    \\
3	&	2451229.43649	&	0.63931	&	$-$83.80	&	88.20	    &	$-$14.79	\\
4	&	2451274.48181	&	0.67440	&	$-$100.02	&	92.98	    &	$-$22.17	\\
5	&	2451275.32141	&	0.10375	&	   68.08	&	$-$71.92	&	$-$16.07	\\
6	&	2451507.50750	&	0.83810	&	$-$97.01	&	96.99	    &	$-$13.32	\\
7	&	2451508.51870	&	0.35521	&	   88.66	&	$-$91.34	&	$-$7.65	    \\
8	&	2451589.50759	&	0.77097	&	$-$113.79	&	112.21	    &	$-$6.71	    \\
9	&	2451593.46730	&	0.79587	&	$-$104.38	&	117.62	    &	$-$6.32	    \\
10	&	2451594.48610	&	0.31686	&	   98.18	&	$-$105.82	&	$-$9.76	    \\
11	&	2451621.40710	&	0.08360	&	   48.85	&	$-$52.15	&	$-$7.32	    \\
12	&	2451625.45629	&	0.15426	&	   89.98	&	$-$94.02	&	$-$6.22	    \\
13	&	2451626.46521	&	0.67020	&	$-$85.20	&	100.80	    &	$-$6.42	    \\
14	&	2451894.43749	&	0.70474	&	$-$106.14	&	113.86	    &	$-$5.44	    \\
15	&	2451896.54610	&	0.78303	&	$-$112.22	&	100.78	    &	$-$6.52	    \\
16	&	2451924.49839	&	0.07716	&	   49.17	&	$-$50.83	&	$-$0.11	    \\
17	&	2451924.58280	&	0.12032	&	   81.00	&	$-$83.00	&	$-$8.28	    \\
18	&	2451929.56758	&	0.66942	&	$-$98.52	&	99.48	    &	$-$1.78	    \\
19	&	2451930.61973	&	0.20746	&	  107.88	&	$-$110.12	&	0.62	    \\
20	&	2451931.40119	&	0.60708	&	$-$74.21	&	67.79	    &	$-$2.46	    \\
21	&	2451983.44518	&	0.22112	&	  120.32	&	$-$105.67	&	$-$4.60	    \\
22	&	2452042.32271	&	0.32478	&	   96.00	&	$-$98.00	&	$-$4.24	    \\
23	&	2452043.34456	&	0.85221	&	$-$91.79	&	92.21	    &	$-$8.03	    \\
24	&	2452272.60939	&	0.09270	&	   63.84	&	$-$67.16	&	$-$15.71	\\
25	&	2452299.42754	&	0.80686	&	$-$98.59	&	98.41	    &	$-$16.91	\\
26	&	2452300.49141	&	0.35089	&	   92.78	&	$-$89.22	&	$-$11.54	\\
27	&	2452303.52260	&	0.90097	&	$-$62.78	&	65.22	    &	$-$15.08	\\
28	&	2452331.44010	&	0.17730	&	  105.87	&	$-$101.13	&	$-$10.22	\\
29	&	2452331.46559	&	0.32034	&	  100.83	&	$-$114.17	&	$-$14.26	\\
30	&	2452331.48858	&	0.20210	&	  105.79	&	$-$114.21	&	$-$14.30	\\
31	&	2452332.49859	&	0.71859	&	$-$108.59	&	109.41	    &	$-$15.68	\\
32	&	2452651.38043	&	0.78705	&	$-$108.89	&	112.71	    &	$-$20.13	\\
33	&	2452652.38195	&	0.29920	&	   107.32	&	$-$111.81	&	$-$17.64	\\
34	&	2452653.35039	&	0.79444	&	$-$108.26	&	110.74	    &	$-$16.09	\\
35	&	2452689.25159	&	0.15344	&	   97.73	&	$-$95.27	&	$-$16.03	\\
36	&	2452689.27739	&	0.16664	&	    92.69	&	$-$103.31	&	$-$21.07	\\
37	&	2452690.30811	&	0.69372	&	$-$99.77	&	113.23	&	$-$15.53	\\
38	&	2452690.32369	&	0.70680	&	$-$104.81	&	114.32	&	$-$12.57	\\
39	&	2453345.49281	&	0.73933	&	$-$112.97	&	113.03	&	$-$18.40	\\
40	&	2453387.57822	&	0.26079	&	117.13	&	$-$114.87	&	$-$17.32	\\
41	&	2453388.54775	&	0.75659	&	$-$115.32	&	113.68	&	$-$25.77	\\
    \hline
    \end{tabular}%
  \label{table_RVs}%
\end{table}%

\begin{table}
    \centering
      \scriptsize

        \caption{Spectroscopic orbit parameters with two different data set.}

    \begin{tabular}{lcc}
         \hline         
         & Direct Measurements & V$_\gamma$ Corrected\\
         \cline{2-3}
         Parameter & \multicolumn{2}{c}{Value} \\
         \hline
         $P$ (d) &  \multicolumn{2}{c}{1.955509 (fixed)} \\
         $T_\mathrm{0}$ ($-2400000$) &  $51224.62848 \pm 0.50149$ & $51224.62942 \pm 0.45640$ \\
         $K_\mathrm{Aa}$ (km\,s$^{-1}$) & $114.46 \pm 0.88$ & $114.31 \pm 0.80$ \\
         $K_\mathrm{Ab}$ (km\,s$^{-1}$) & $111.91 \pm 0.88$ & $112.07 \pm 0.80$ \\
         $e$  & $0.0021 \pm 0.0014$ & $0.0021 \pm 0.0013$\\
         $w~(^\circ)$ & $155 \pm 81$ & $155 \pm 74$\\ 
         $V_\gamma$ (km\,s$^{-1}$) & $-16.35 \pm 0.52$ & $-0.25 \pm 0.48$ \\
         $m_\mathrm{Aa}\sin{^3{i}}$ ($M_\odot)$      & $1.188 \pm 0.009$  & $1.187 \pm 0.008$  \\
         $m_\mathrm{Ab}\sin{^3{i}}$ ($M_\odot)$      & $1.162 \pm 0.009$  & $1.164 \pm 0.008$ \\
         \emph{q} ($M_\mathrm{Ab}/M_\mathrm{Aa}$)     & $0.978 \pm 0.015 $ &  $0.981 \pm 0.013$ \\
         $a~\sin{i}_\mathrm{A}$  ($R_\odot$)        & $8.696 \pm 0.067 $ &  $8.696 \pm 0.061 $ \\
         rms (km\,s$^{-1}$) & 4.72 & 4.30\\
         \hline
    \end{tabular}
    \label{tab:spec_orbs_a}
\end{table}

\newpage
\subsection{Close Binary System - Aab}

PHOEBE v2.4.13, renowned for its advanced capabilities in modeling eclipsing binary systems with intricate physical processes and precise parameter estimation, was employed to determine the parameter estimates for the close binary system (Aab) \citep{phoebe1, phoebe2, phoebe3, phoebe4, phoebe5}. Previous studies had not found evidence for mass transfer nor the current study's O--C analysis, so the components were assumed to be `detached' for the analysis. Conjunction time $T_0$, orbital period $P$, and the temperature of the primary component (Aa) were fixed. The primary star temperature was adopted from \cite{Marrese2004} as \ppnumber{6500} K. The following parameters were adjusted: mass ratio ($q$), eccentricity ($e$), the argument of periapsis ($w$), semi-major axis ($a$), systemic velocity ($V_\gamma$), orbital inclination ($i$), temperature ratio for the primary (Aa) and the secondary (Ab) components ($T_2/T_1$), the radii of the primary (Aa) and the secondary (Ab) components ($R_{1,2}$), and monochromatic luminosity for the primary (Aa) and tertiary (B) components ($L_1$ and $L_3$).

It should be noted that the temperatures of the components cannot be measured directly from the analysis of light curve (LC) and RV data. We simply selected the primary (Aa) component's temperature for reference. The determination of the temperatures of the components is explained in \S3.4. The monochromatic luminosity for the tertiary (B) component was analyzed as a fraction of total light. Limb darkening coefficients were calculated for the primary (Aa) and the secondary component (Ab) internally by \texttt{PHOEBE} via using \citet{Castelli2004} stellar atmosphere models. Considering the mass and temperature values for the primary (Aa) and the secondary (Ab) components from the previous studies, the gravity-darkening coefficients and bolometric albedos were selected as 0.32 and 0.5 for the primary (Aa) and the secondary (Ab) components, respectively \citep{Lucy1967, Rucinski1969}. The initial analysis used \citet{Nelder} optimization. Our modeling used the \textit{TESS} photometric data to calculate the fundamental parameters of the A component of CN Lyn since it's more precise and reliable than \textit{Hipparcos} photometric data ($H_{\rm p}$), which is used to solely calculate the light contribution of components. 

\begin{figure}
  \centering
  \includegraphics[width=0.9\linewidth]{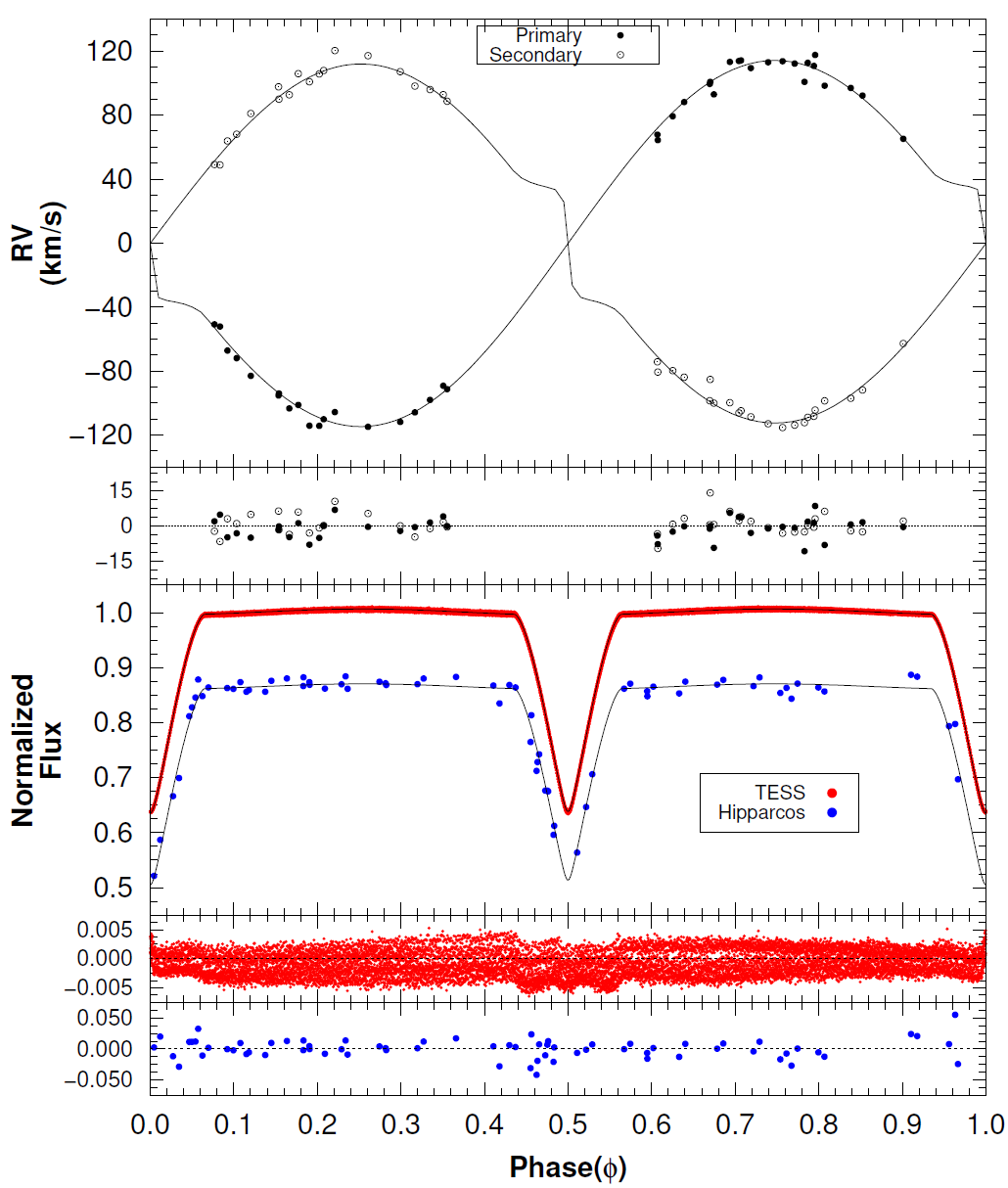}
   \caption{Observed RVs with best fitting radial velocity curves and photometric data with LC modelling. Filled and empty circles represent the RVs of the primary and secondary components of Aab, respectively. In the LC panel, red and blue dots represent the \textit{TESS} and \textit{Hipparcos} data, respectively, and the black curves are the best LC model for each photometric data set.}
   \label{cnlyn_rvlc}
\end{figure}

\begin{figure*}
    \centering
    \includegraphics[width=0.9\textwidth]{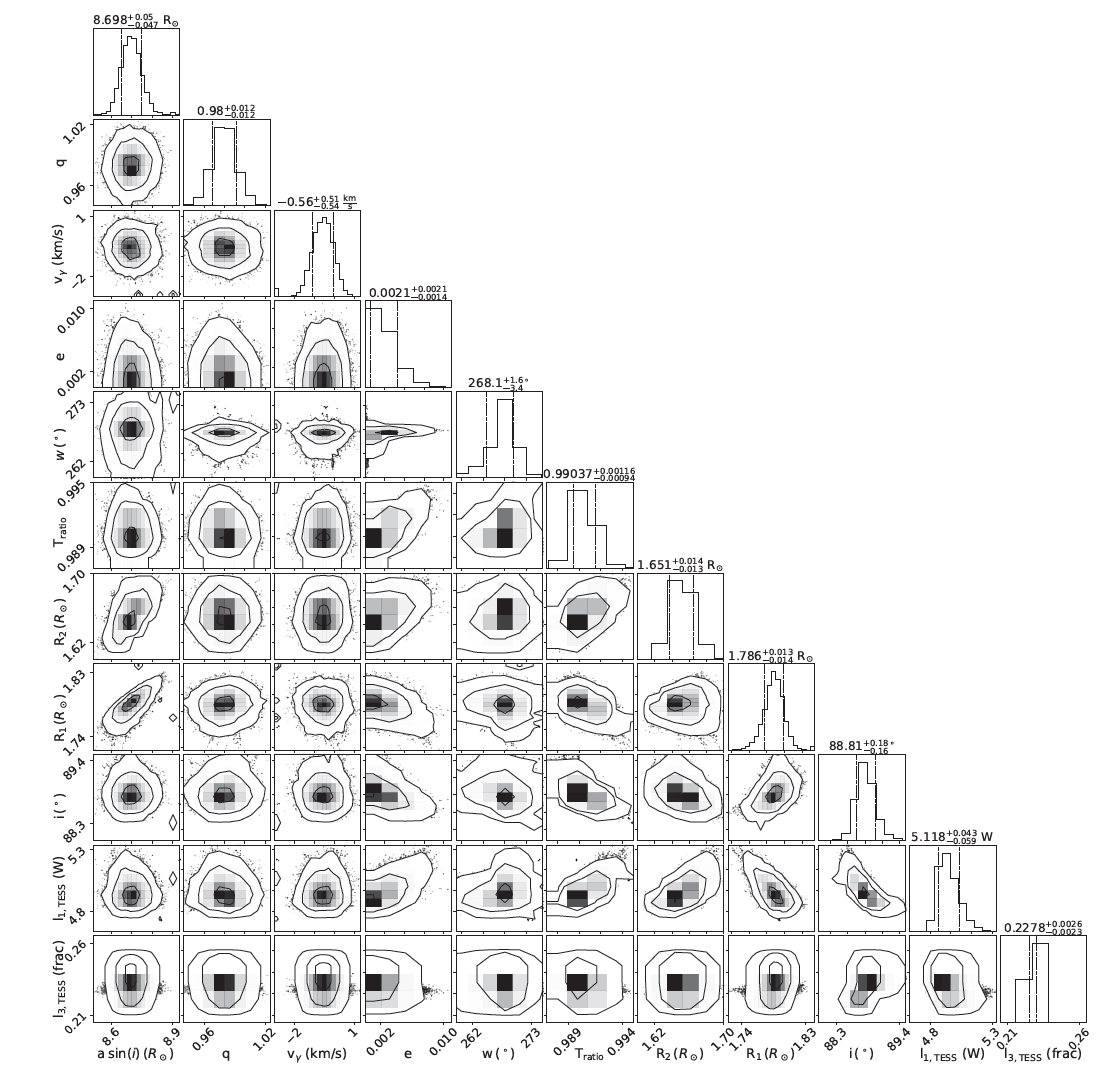}
    \caption{A corner plot of the posteriors for the fundamental parameters of Aab components.}
    \label{MCMC}
\end{figure*}

To calculate uncertainties, we performed a Markov Chain Monte Carlo (MCMC) optimization using this built-in feature of \texttt{PHOEBE} \citep{mcmc}. We performed the analysis at T\"{U}B{\.{I}}TAK ULAKB{\.{I}}M, the High Performance and Grid Computing Center (TRUBA) using 128 walkers with \ppnumber{1000} iterations. The LC and spectroscopic orbit models are presented in Figure~\ref{cnlyn_rvlc} and the fundamental parameters and heuristic errors for the close binary system (A) of CN Lyn are presented in Figure~\ref{MCMC} and listed in Table~\ref{shortfund}, respectively.

\begin{table}
	\setlength{\tabcolsep}{2pt}
	\renewcommand{\arraystretch}{1.4}
\centering
  \scriptsize
\caption{Parameters determined for Aab from the analysis of RV and LC data.} 
\label{tab:absolutepar}
\begin{tabular}{lccc}\hline
Parameter                       & Symbol                                & Primary (Aa)                      & Secondary (Ab) \\
\hline
Separation ($R_\odot$)          & $a\sin i$                             & \multicolumn{2}{c}{$8.698^{+0.050}_{-0.047}$} \\
Mass ratio                      & \emph{q}                              & \multicolumn{2}{c}{$0.980^{+0.012}_{-0.012}$} \\
Eccentricity                    & \emph{e}                              & \multicolumn{2}{c}{$0.0021^{+0.0021}_{-0.0014}$} \\
Argument of perigee ($^\circ$)  & \emph{w}                              & \multicolumn{2}{c}{$268.1^{+1.6}_{-3.4}$} \\
Orbital inclination ($^\circ$)  & \emph{i}                              & \multicolumn{2}{c}{$88.81^{+0.18}_{-0.16}$} \\
Temperature ratio               & $T_\mathrm{eff,b}/T_\mathrm{eff,a}$   & \multicolumn{2}{c}{$0.9904^{+0.0012}_{-0.0009}$} \\
Mass ($M_\odot$)                & \emph{M}                              & $1.166^{+0.013}_{-0.012}$         & $1.143^{+0.013}_{-0.012}$ \\
Radius ($R_\odot$)              & \emph{R}                              & $1.786^{+0.013}_{-0.014}$         & $1.651^{+0.014}_{-0.013}$ \\
Surface gravity (cgs)           & $\log g$                              & $4.001^{+0.012}_{-0.011}$         & $4.061^{+0.011}_{-0.012}$ \\
Light ratio (\textit{TESS})     & $l/l_\mathrm{total}$                  & $0.4073^{+0.0034}_{-0.0047}$      & $0.3649^{+0.0047}_{-0.0034}$ \\
\hline
\label{shortfund}
\end{tabular}
\end{table}

\vspace{10pt}
\subsection{Spectral Disentangling} 

The spectra of multiple systems can be challenging to interpret. There are ways, such as making a grid of synthetic spectra for each component, then combining it with other ones using light contributions for each component, and finally matching it with the original spectra to determine the temperature and metallicity of the components \citep{Yucel2022}. However, the best way to determine the temperature and to obtain chemical abundances of a component in a multiple system is to disentangle the spectrum of each component in the system. Thus, one can analyze each spectrum and determine the temperature and chemical properties of each component independently.

\begin{figure*}
    \centering
    \includegraphics[width=\linewidth]{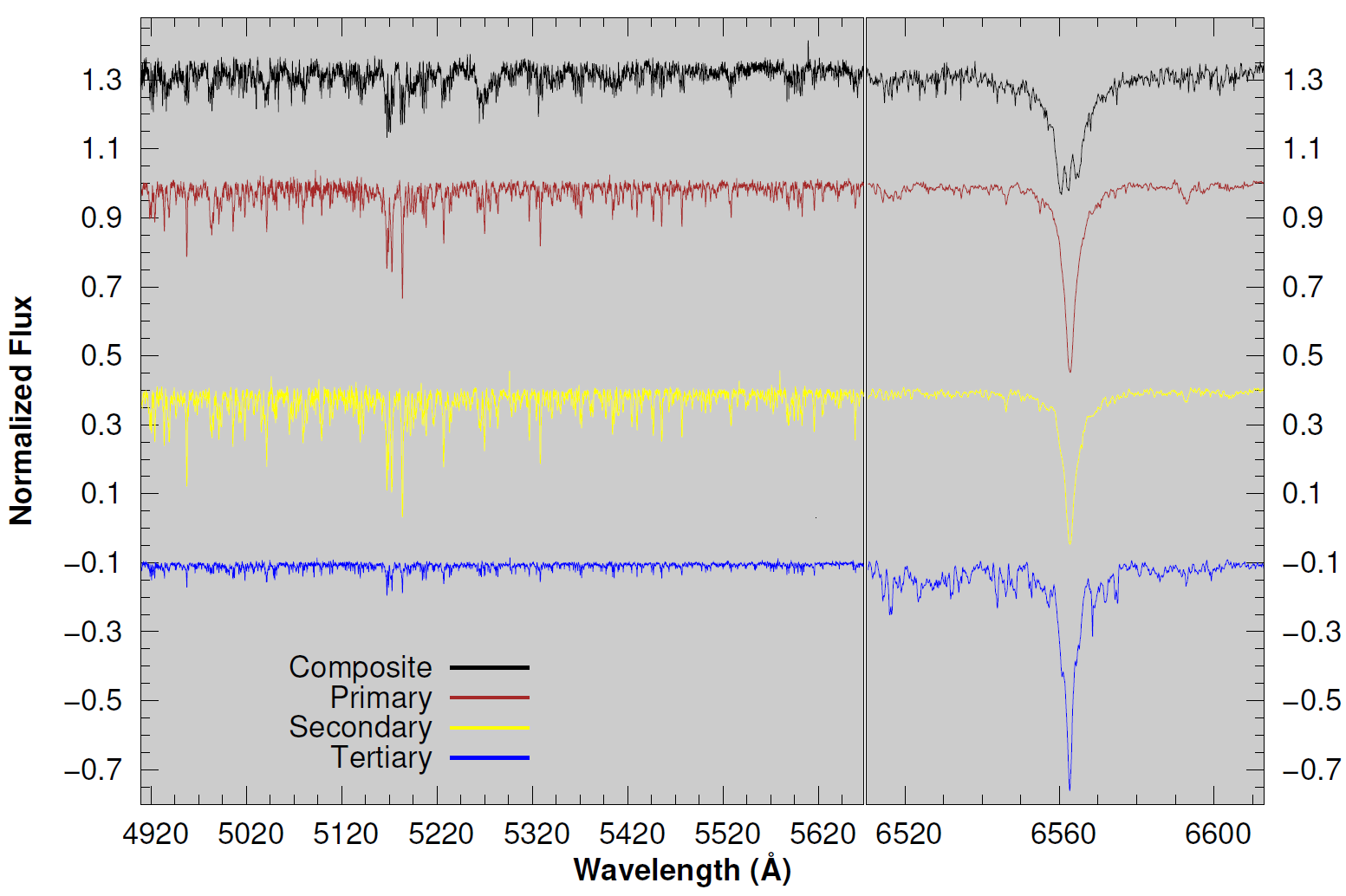}
    \caption{Disentangled component spectra are shown together with observed composite spectrum.}
    \label{disentangle}
\end{figure*}

Before starting the disentanglement procedure, there is a challenge for the current study. Our spectroscopic data contains two different spectroscopic data sets. Each one has a different resolution and, therefore, a different dispersion in terms of \AA/pixel. The data need to have the same step size to combine the data for analysis. To resolve this problem, we implemented an algorithm that, basically, reads spectra from two different sets and makes step size the same. We interpolated the ELODIE data to the same scale as the ASIAGO data. In terms of data quality, there is a loss in resolution from higher resolution data, in this case, the ELODIE data, but in the end, we gain $S/N$, which yields more accurate results. The final data sets had \ppnumber{2048} data points covering a wavelength range of around $125$ \AA\, for each spectrum. After this procedure, we used the Fourier disentangle code \texttt{KOREL} \citep{Korel1, Korel2}\footnote{\url{https://stel.asu.cas.cz/vo-korel}} to obtain the individual spectrum of each component of the system (CN Lyn AabB). The disentangled component spectra are shown in Figure~\ref{disentangle}.


\subsection{Determination of Temperatures and Chemical Abundances}

\begin{figure*}[h]
    \centering
    \includegraphics[width=.7\linewidth]{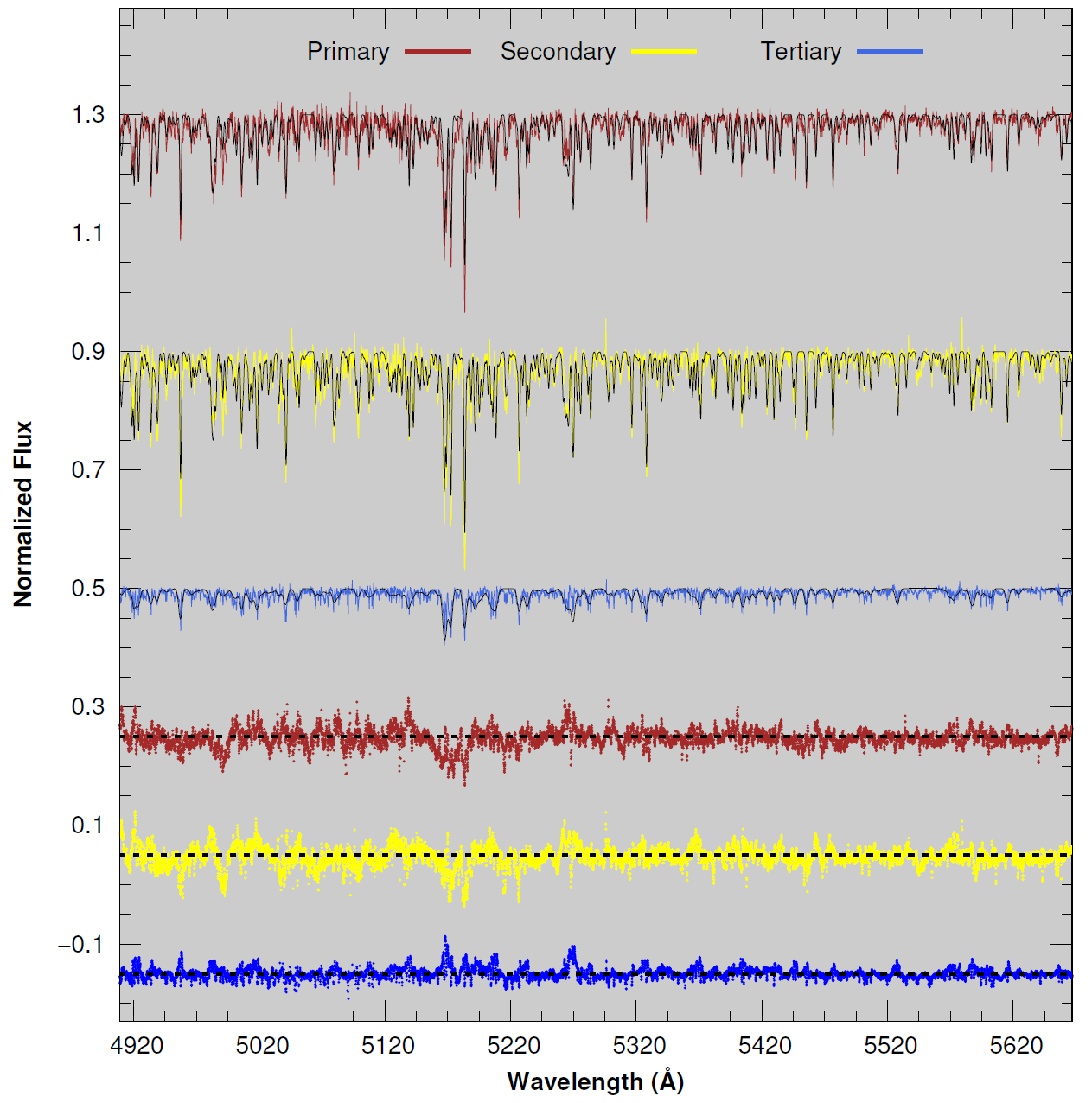}
     \includegraphics[width=.7\linewidth]{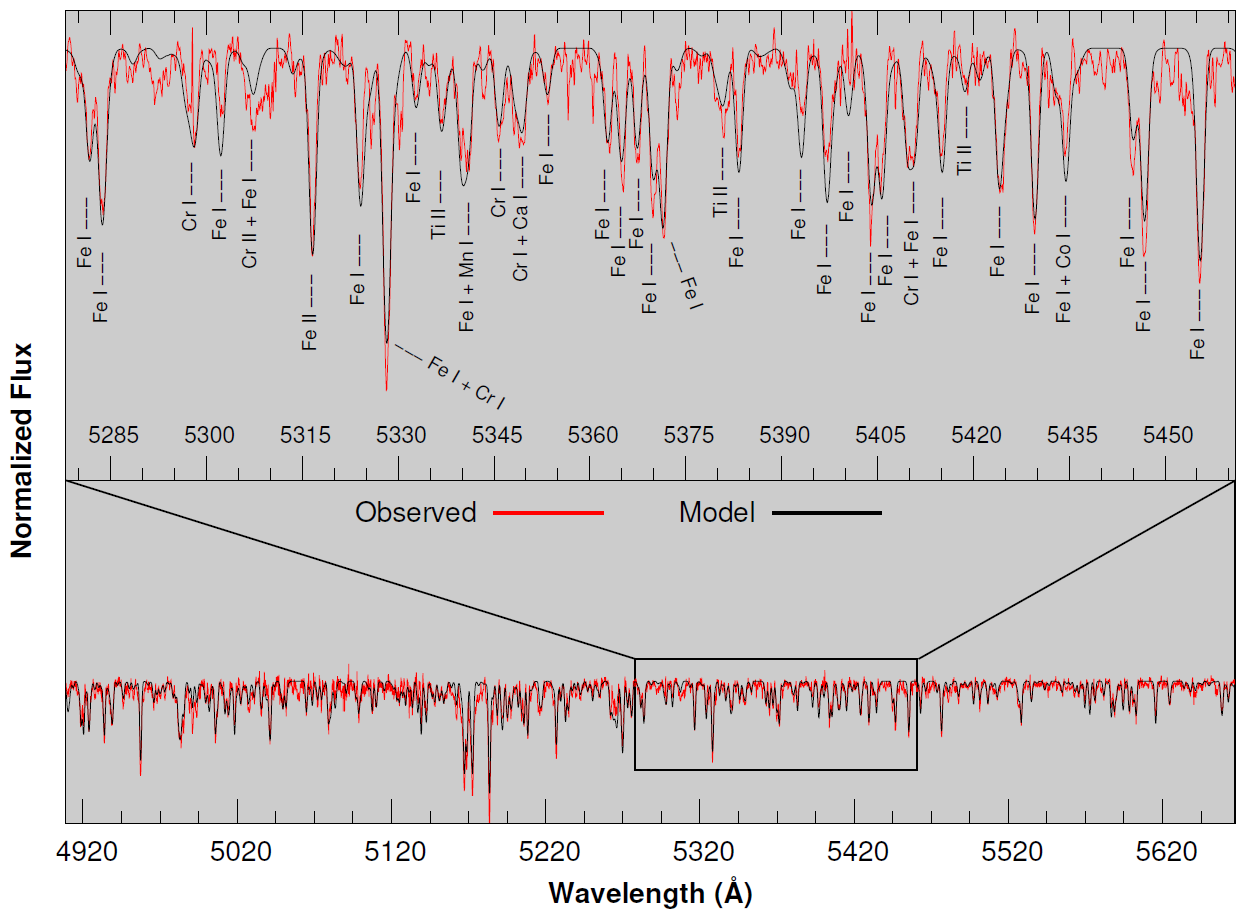}
    \caption{\textit{Upper panel}: Each component's disentangled spectra with SP\_Ace models. Lower parts in the panel show the residuals between the disentangled spectra and SP\_Ace models. \textit{Lower panel}: Highlighted region from the disentangled spectra of the primary component with SP\_Ace model shows each element used in chemical abundance calculations.}
    \label{fig:sp_ace}
\end{figure*}

\begin{table*}[t]
    \centering
    \caption{Results of analysis of SP\_Ace for each component. The metal abundance ([M/H]) was calculated according to the formula of \cite{Salaris1993}.}
    \begin{tabular}{lccccccc}
    \hline
        Parameter & Symbol & \multicolumn{2}{c}{Primary} & \multicolumn{2}{c}{Secondary} & \multicolumn{2}{c}{Tertiary} \\
        \hline
        Temperature & $T_\mathrm{eff}$ (K)& \multicolumn{2}{c}{$6411^{+67}_{-40}$} & \multicolumn{2}{c}{$6406^{+40}_{-94}$} & \multicolumn{2}{c}{$6238^{+52}_{-49}$}\\
        Signal/Noise & $S/N$ & \multicolumn{2}{c}{82} & \multicolumn{2}{c}{75} & \multicolumn{2}{c}{131}\\
        Chi-square & $\chi^2$ & \multicolumn{2}{c}{2.10} & \multicolumn{2}{c}{2.24} &  \multicolumn{2}{c}{2.19} \\
           \hline
\multicolumn{2}{c}{Abundance (dex)} & Number of Lines & Primary & Number of Lines & Secondary & Number of Lines & Tertiary \\
        \hline
 \multicolumn{2}{c}{[M/H]} & 719 & $-0.68^{+0.04}_{-0.01}$ & 778 & $-0.47^{+0.01}_{-0.06}$ & 466 &$-1.47^{+0.05}_{-0.05}$\\
 \multicolumn{2}{c}{[Fe/H]}  	&	525	& $-0.78^{+0.02}_{-0.02}$ & 613 & $-0.55^{+0.03}_{-0.02}$ & 349 & $-1.83^{+0.09}_{-0.11}$ \\

 \multicolumn{2}{c}{[C/H]}   	&	15	& $-0.62^{+0.17}_{-0.18}$ & 16 & $-0.04^{+0.09}_{-0.16}$ & --- & ---\\

 \multicolumn{2}{c}{[Na/H]}  	& 8 & $-0.24^{+0.12}_{-0.12}$ & 4 &	 --- & --- & ---\\

 \multicolumn{2}{c}{[Mg/H]}  	& 2 & $-0.68^{+0.10}_{-0.13}$ & 2 & $-0.52^{+0.17}_{-0.18}$ & --- & ---	\\

 \multicolumn{2}{c}{[Al/H]}      & 1 & --- & 2 & --- & --- & ---\\

 \multicolumn{2}{c}{[Si/H]}  	& 23 & $-0.52^{+0.08}_{-0.09}$ & 30 &  $-0.41^{+0.10}_{-0.13}$ & --- &	---\\

 \multicolumn{2}{c}{[Ca/H]}  	& 41 & $-0.62^{+0.06}_{-0.04}$ & 21 &	 $-0.41^{+0.10}_{-0.13}$ & 30 &	 $-1.35^{+0.05}_{-0.20}$	\\

 \multicolumn{2}{c}{[Sc/H]}  	& 23 & $-0.73^{+0.08}_{-0.04}$ & 16 & $-0.46^{+0.05}_{-0.10}$ & 21 & $-1.03^{+0.10}_{-0.34}$\\

 \multicolumn{2}{c}{[Ti/H]}  	& 128 & $-0.70^{+0.06}_{-0.04}$ & 112 & $-0.66^{+0.05}_{-0.07}$ & 87 & $-1.17^{+0.10}_{-0.11}$\\

 \multicolumn{2}{c}{[V/H]}   	& 6 & $-0.81^{+0.44}_{-0.43}$ & 8 &	 --- & --- & ---                   \\

 \multicolumn{2}{c}{[Cr/H]}  	& 109 & $-0.88^{+0.04}_{-0.05}$ & 127 & $-0.57^{+0.04}_{-0.05}$ & 48 &	 $-1.44^{+0.04}_{-0.04}$\\

\multicolumn{2}{c}{[Mn/H]}  	& 37 & $-0.99^{+0.18}_{-0.17}$  & 62 & $-0.89^{+0.19}_{-0.20}$ & --- & ---\\

\multicolumn{2}{c}{[Co/H]}      & 32 & --- & 43 & --- & 5 &	 $-1.75^{+0.41}_{-0.40}$\\

\multicolumn{2}{c}{[Ni/H]}  	& 108 & $-0.77^{+0.05}_{-0.03}$ & 119 & $-0.40^{+0.04}_{-0.05}$ & 91 & $-1.33^{+0.24}_{-0.07}$\\

\multicolumn{2}{c}{[Cu/H]}  	& 4 &  $-0.63^{+0.22}_{-0.27}$ & 6 & $-0.39^{+0.21}_{-0.26}$ & --- & ---\\

\multicolumn{2}{c}{[Y/H]}  	    &	16	&	 $-0.56^{+0.16}_{-0.13}$ & 16 &	 --- & --- & --- \\

\multicolumn{2}{c}{[Zr/H]}  	&	5	&	 $-0.55^{+0.38}_{-0.39}$ & 5 & $-0.03^{+0.21}_{-0.40}$ & --- & ---\\

\multicolumn{2}{c}{[Ba/H]}  	& 1 & $-0.25^{+0.26}_{-0.33}$ & 1 & $-0.29^{+0.44}_{-0.43}$ & --- & ---	\\

\multicolumn{2}{c}{[Ce/H]}  	& 9 & $-0.59^{+0.17}_{-0.17}$ & 14 & $-0.39^{+0.22}_{-0.31}$ & 2 &	 $-1.41^{+0.59}_{-0.61}$\\

\multicolumn{2}{c}{[Nd/H]}  	& 15 & $-1.05^{+0.67}_{-0.67}$ &	22 &  $-0.28^{+0.19}_{-0.46}$ &  --- & --- \\
        \hline
    \end{tabular}
    \label{tab:chem_results}
\end{table*}

\begin{figure*}[t]
    \centering
    \includegraphics[width=\linewidth]{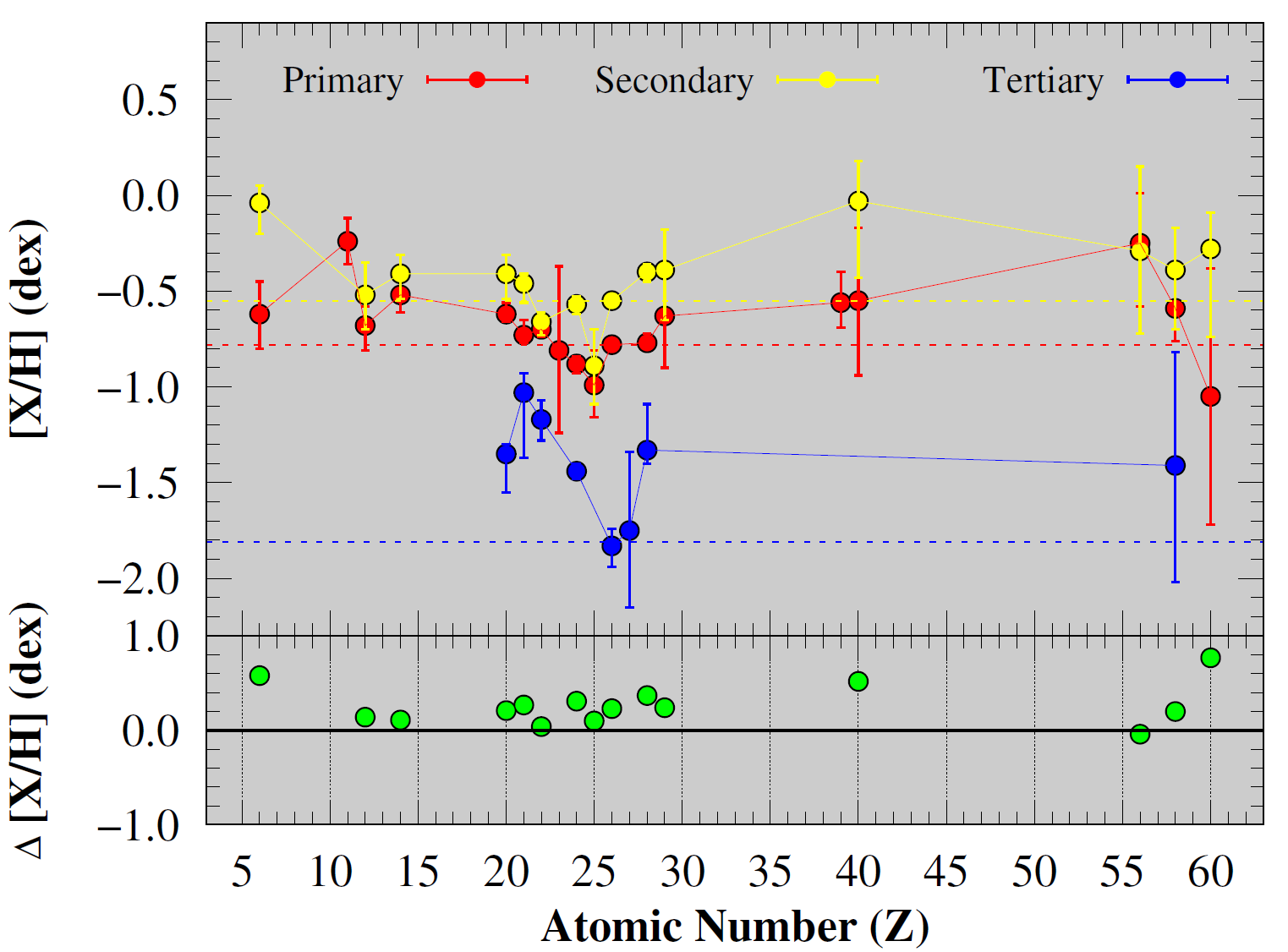}
    \caption{\textit{Upper panel}: The distribution of chemical abundances of each component of CN Lyn versus atomic number. Red, yellow, and blue colors represent the primary, secondary, and tertiary components, respectively. Dashed lines represent [Fe/H] value of each component. \textit{Lower panel}: Differences of chemical abundances $\left([X/H]_\mathrm{Ab}-[X/H]_\mathrm{Aa}\right)$ in components of the A system according to atomic number.}
    \label{fig:atomic}
\end{figure*}

Precise temperature and chemical abundance determination from a spectrum requires high $S/N$ and high-resolution data \citep{Jofre2019}. Although our data possess sufficient resolution for spectral analysis to calculate temperature and chemical abundances from absorption lines, the signal-to-noise ratio ($S/N$) is inadequate and so an alternative method needed to be considered.

Consequently, we utilized an automated code to calculate temperature and chemical abundances. Specifically, we employed version 1.4 of the \texttt{SP\_Ace} code \citep{space1, space2} to determine the temperature and chemical abundances of each component. \texttt{SP\_Ace} calculates stellar atmosphere parameters ($T_{\mathrm{eff}}$, log \emph{g}, [Fe/H], and chemical abundances) by constructing many spectrum models with different temperatures, surface gravities, and chemical abundances. It then finds the best match with the observed spectrum via a $\chi^2$ optimization routine. \texttt{SP\_Ace} uses updated \texttt{ATLAS9} \citep{Castelli2003} grids\footnote{\url{https://wwwuser.oats.inaf.it/fiorella.castelli/grids.html}} and \texttt{SPECTRUM} \citep{Gray1994} code to produce synthetic spectra. More details can be found in the original studies \citep{space1, space2}. Our first analysis indicated that there is a difference in chemical abundances between the primary (Aa) and the secondary (Ab) components. Therefore, we analyzed different wavelength regions to assess the impact of the number of lines, but the results remained consistent. We have specifically selected [Fe/H] lines since it has the most abundant chemical line in the samples. It should be noted that in our analysis, surface gravity values for the primary (Aa) and the secondary (Ab) components have been fixed with values calculated from RV and LC analysis since these values are more precise than the surface gravity calculated from spectroscopic analysis. The elemental abundances estimated for the components in the CN Lyn system are presented in Table \ref{tab:chem_results}.  In Figure \ref{fig:sp_ace} each component's disentangled spectra with SP\_Ace models, its residuals, and, also, a highlighted region from the disentangled spectra of the primary (Aa) component with each element labeled is given. A comparison of the chemical abundances in each component is given in Figure \ref{fig:atomic}. As it can be seen in Figure \ref{fig:atomic}, the Ab component is more metal-rich than the Aa component in terms of almost all chemical abundances. The difference in [Fe/H] between the Ab component and the Aa component is calculated as $\Delta$[Fe/H]=+0.23\,dex. The comparison between the Ab component and the B component in terms of individual chemical abundances is in the range of $+0.57 \leq \Delta\,{\rm [X/H] (dex)} \leq +1.28$ and the difference in iron abundances is $\Delta$[Fe/H]=+1.28\,dex, as well.

\subsection{Fundamental Parameters of the Components of CN Lyn}

Combining the re-calculated precise radial velocities with accurate photometric \textit{TESS} data allowed calculation of the fundamental parameters of the analyzed systems with a precision better than $\sim$1.5 and $\sim$1\% for mass and radius, respectively, in the primary (Aa) and the secondary (Ab) components. 

It should be noted that the results for temperature and chemical abundance calculations of \texttt{SP\_Ace} are reliable, however, uncertainties should be approached carefully. Detailed spectroscopic studies of single stars with high $S/N$ and high-resolution data indicate that uncertainties in the calculation of temperature and chemical abundances cannot be better than 50 K and 0.1 dex, therefore, considering the quality of our data, we chose the uncertainties in our calculations as 100 K and 0.15 dex for temperature and chemical abundances, respectively. The radius of the third component (B) was calculated using light contributions.

Calculating the temperature of each component enables us to determine their spectral types and intrinsic colors. Bolometric and absolute magnitudes have been derived according to the method described by \cite{Eker2020, Eker2021, Eker2022}. Thus, the photometric distance of the system has been carried out using the distance modulus. The calculated photometric distance is in excellent agreement with \textit{Gaia} trigonometric parallax (see Table~\ref{table:fund}), which implies that our solutions are on point.

\begin{table*}[p!]
	\setlength{\tabcolsep}{5pt}
	\renewcommand{\arraystretch}{1.2}
\centering
\caption{Multi-stellar parameters and heuristic errors of CN~Lyn.} \label{tab:parameters}
\begin{tabular}{lcccc}\hline
Parameter & Symbol  & Primary & Secondary & Tertiary \\
\hline
Equatorial coordinate (Sexagesimal)             & $(\alpha, \delta)_{\rm J2000}$ & \multicolumn{3}{c}{08:01:37.20, +38:44:58.41} \\
Galactic coordinate (Decimal)                   & $(l, b)_{\rm J2000}$ & \multicolumn{3}{c}{181.759605, +29.646775} \\
Ephemerides time (d)                            & \emph{T}$_{\rm 0}$    & \multicolumn{2}{c}{$2459947.800935_{-0.00001}^{+0.00001}$}      &   $2448783_{-26}^{+26}$  \\ 
Orbital period (d)                              & \emph{P}              & \multicolumn{2}{c}{$1.955509_{-0.00001}^{+0.000001}$} & $3130_{-78}^{+78}$              \\ 
Separation ($R_\odot$)                          & \emph{a}              & \multicolumn{2}{c}{$8.700^{+0.051}_{-0.048}$} & $1327_{-176}^{+175}$  \\ 
Combined visual magnitude                       & \emph{V}              & \multicolumn{3}{c}{$9.06_{-0.02}^{+0.02}$}                              \\
Combined visual magnitude$^1$                   & \emph{TESS}           & \multicolumn{3}{c}{$8.528_{-0.006}^{+0.006}$}                           \\
Combined color index (mag)                     & $B-V$                 & \multicolumn{3}{c}{$0.43_{-0.04}^{+0.04}$}                              \\
Color excess (mag)                              & $E(B-V)$              & \multicolumn{3}{c}{$0.05_{-0.05}^{+0.05}$}                                                     \\
Systemic velocity (km\,s$^{-1}$)                & $V_{\gamma}$          & \multicolumn{3}{c}{$-14.9_{-0.4}^{+0.4}$}                         \\

Orbital inclination ($^{\circ}$)                & \emph{i}              & \multicolumn{2}{c}{$88.81_{-0.16}^{+0.18}$}                     & $26.7_{-2.1}^{+2.1}$ \\
Mass ratio                                      & \emph{q}              & \multicolumn{2}{c}{$0.980_{-0.012}^{+0.012}$}                      & $0.37_{-0.09}^{+0.09}$ \\
Eccentricity                                    & \emph{e}              & \multicolumn{2}{c}{$0.0021_{-0.0014}^{+0.0021}$}                & $0.55_{-0.14}^{+0.14}$ \\
Argument of perigee (rad)                       & \emph{w}              & \multicolumn{2}{c}{$4.677_{-0.059}^{+0.028}$}                & $3.421_{-0.140}^{+0.139}$ \\
Spectral type                                   & Sp                    & F6 V                               & F6 V                       & F7-8 V-IV \\
Metallicity (dex)                               & [Fe/H]                & $-0.78_{-0.02}^{+0.02}$            & $-0.55_{-0.02}^{+0.03}$ & $-1.83_{-0.11}^{+0.09}$ \\
Mass ($M_\odot$)                                & \emph{M}              & $1.166_{-0.012}^{+0.013}$           & $1.143_{-0.012}^{+0.013}$   & $0.85_{-0.23}^{+0.23}$\\
Radius ($R_\odot$)                              & \emph{R}              & $1.786_{-0.014}^{+0.013}$           & $1.651_{-0.013}^{+0.014}$   &  $1.436_{-0.023}^{+0.026}$ \\
Surface gravity (cgs)                           & $\log g$              & $4.001_{-0.011}^{+0.012}$           & $4.061_{-0.012}^{+0.011}$   & $4.053_{-0.153}^{+0.120}$ \\
Age (Gyr)                                       & \emph{t}              & \multicolumn{2}{c}{$3.89_{-0.10}^{+0.10}$}  & $12.50_{-2.50}^{+2.50}$                      \\   
Light ratio (\textit{TESS})                     & $l/l_{\rm{total}}$    & $0.4073^{+0.0034}_{-0.0047}$         & $0.3649^{+0.0047}_{-0.0034}$   & $0.2278_{-0.0023}^{+0.0026}$ \\
Temperature (K)                                 & $T_{\rm eff}$         & $6411_{-100}^{+100}$            & $6406_{-100}^{+100}$    & $6238_{-100}^{+100}$ \\
Luminosity ($L_\odot$)                          & $\log$ \emph{L}       & $0.700_{-0.034}^{+0.033}$           & $0.619_{-0.035}^{+0.034}$   & $0.448_{-0.042}^{+0.043}$ \\
Individual \emph{TESS} magnitude (mag)                  &  \emph{TESS$_{1,2,3}$}   & $9.503_{-0.018}^{+0.018}$           & $9.629_{-0.019}^{+0.019}$ & $10.134_{-0.012}^{+0.013}$ \\
Individual visual magnitude (mag)                    &  \emph{V$_{1,2,3}$}    & $10.015_{-0.028}^{+0.028}$           & $10.135_{-0.028}^{+0.028}$   & $10.646_{-0.034}^{+0.034}$ \\
Bolometric magnitude (mag)$^2$                           & $M_{\rm bol}$         & $2.990_{-0.082}^{+0.085}$           & $3.192_{-0.085}^{+0.088}$   &  $3.621_{-0.207}^{+0.105}$ \\
Absolute \textit{TESS} magnitude (mag)$^3$                    & $M_{\rm TESS}$           & $2.786_{-0.028}^{+0.029}$           & $2.906_{-0.029}^{+0.028}$  & $3.423_{-0.036}^{+0.032}$ \\
Absolute visual magnitude (mag)                      & $M_{\rm V}$           & $2.980_{-0.081}^{+0.081}$           & $3.194_{-0.088}^{+0.088}$  & $3.678_{-0.093}^{+0.093}$ \\
Bolometric correction (mag)$^4$     & \emph{BC$_{\rm TESS}$} & $0.204_{-0.111}^{+0.113}$           & $0.406_{-0.114}^{+0.116}$   & $0.198_{-0.239}^{+0.141}$ \\
Bolometric correction (mag)                 &  \emph{BC$_{\rm V}$}   & $0.010_{-0.173}^{+0.176}$           & $0.002_{-0.176}^{+0.173}$   & $-0.057_{-0.300}^{+0.198}$ \\
Computed synchronization velocity (km\,s$^{-1}$)& $v_{\rm synch}$       & $46.2_{-0.4}^{+0.4}$                 & $46.7_{-0.4}^{+0.4}$  & --- \\
Photometric distance (pc)                                   &  \emph{d}              & \multicolumn{3}{c}{$241_{-10}^{+10}$}                                      \\
{\it Gaia} distance (pc)                        & $d_{\,\varpi}$              & \multicolumn{3}{c}{$230.9\pm 3.4$}                                       \\
\hline
\multicolumn{5}{l}{$^1$\cite{Paegert2022},$^2$\cite{Eker2022},$^3$\cite{Bakis2022},$^4$\cite{Eker2023}}
\label{table:fund}
\end{tabular}
\end{table*}

\section{Evolutionary Status} \label{sec:evolution}

We have used version r23.05.1 of Modules for Experiments in Stellar Astrophysics \citep[\texttt{MESA},][]{Paxton2011, Paxton2013, Paxton2015, Paxton2018, Paxton2019, Jermyn2023} for evolution calculations. The evolution of multiple systems is tricky, since orbital elements, period, and eccentricity (if present)  are also changing during evolution because of angular momentum losses in the systems, besides the evolution of the stars in a system. One must consider both in terms of the calculation of evolutionary scenarios of binary systems. The \texttt{binary} module of \texttt{MESA} considers both orbital parameter changes and the natural evolution of stars in binary systems, simultaneously. In the following, evolutionary scenarios of the close binary (Aab) and the distant star (B) have been investigated. 

\subsection{Evolution Analysis of Aab}

Our analysis and current information in the literature indicate that Aab is detached and there is no prior mass transfer since the system has entered the main-sequence. Therefore, the changes, in evolution terms, only have happened in the components themselves, because of nuclear reactions in their core, and in the orbital parameters, period, and eccentricity. As a general rule in the binary evolution calculations, to determine the initial orbital conditions of the Aab component of CN Lyn when the components entered the main-sequence phase and started its evolution, we used a similar approach that has been used widely in literature: a grid search \citep[e.g.][]{Rosales, Soydugan, Yucel2022, Yucel2024}. Grid searches can be computationally demanding and require long computer-running times. Therefore, before making a grid search with different starting orbital elements, a starting point is needed for efficiency. In this regard, we performed evolution calculations that started with a randomly selected initial period and initial eccentricity and finished the evolution when the evolution reached the up-to-date eccentricity value $0.0021_{-0.0014}^{+0.0021}$) of the system, Aab. If the system had a circular orbit, we could not do such an analysis as we would not know reliably when the circularization has been achieved. 

\begin{figure}[!t]
    \centering
    \includegraphics[width=0.98\linewidth]{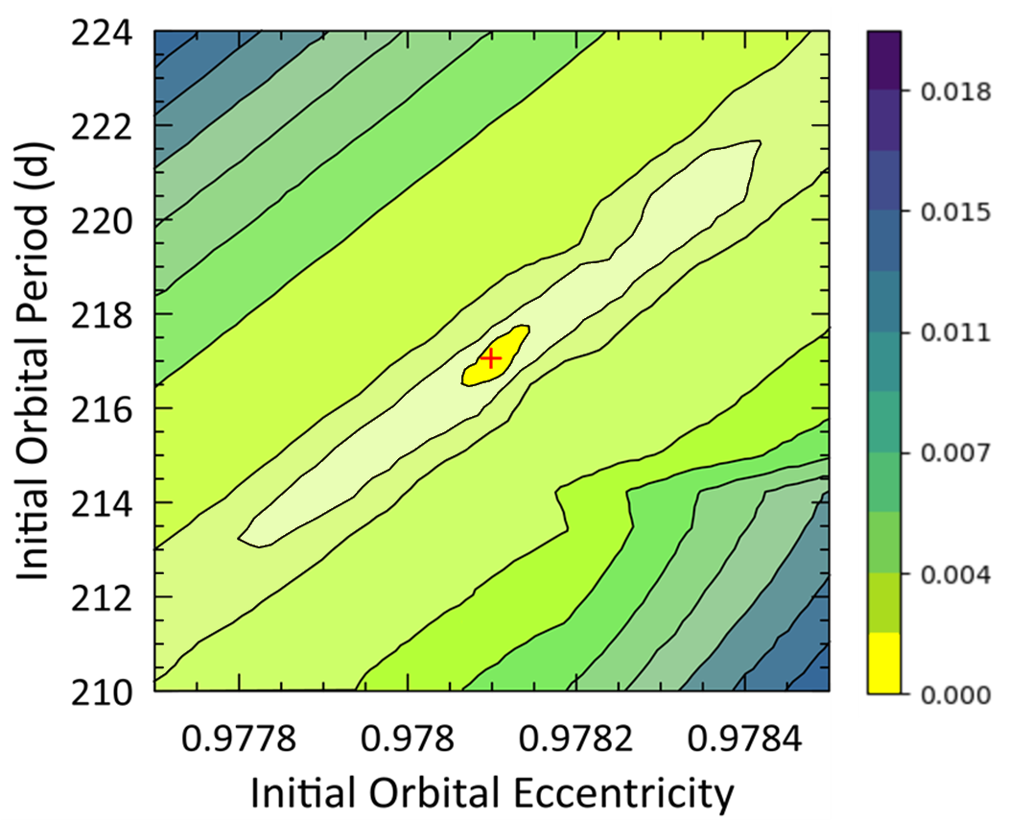}
    \caption{The result of the grid search. The best model is given by red plus in the figure with initial period and initial eccentricity, as 217 day and 0.9781, respectively. The colorbar shows $\chi^2$ values.}
    \label{fig:chi}
\end{figure}

Furthermore, after we determined a starting point, we did a grid search that initial period changes between 200 and 230 days with an interval of 0.5 days and initial eccentricity changes between 0.9775 and 0.9790 with an interval of 0.0001, based on our starting point search. Then, a $\chi^2$ calculation was made using the determined orbital period and eccentricity of the system, calculated radii, and temperature of the components with every model in the grid. According to our calculations the best fitting model (of lowest $\chi^2$, in this case 0.00159) indicated initial orbital parameters for period and eccentricity as 217 days and 0.9781, respectively (given in Figure \ref{fig:chi}). In our calculations, based on both components having convective atmospheres, we included magnetic breaking \citep{magnetic}. For the tidal synchronization, we used the ``Orb\_period'' option, which synchronizes the orbit relevant to the timescale of the orbital period. We also applied tidal circularization, given by \cite{Hurley}. Roche lobe radii in binary systems are computed using the relation given by \cite{Eggleton1983}. Mass transfer rates in Roche lobe overflowing binary systems are determined following the prescription of \cite{Kolb1990}. 

After determining the initial orbital conditions, we conducted an additional evolutionary analysis, extending from the onset of mass transfer until the secondary star reached the Terminal Age Main Sequence (TAMS). This analysis was based on the calculated initial orbital parameters determined above. The mean rates for mass transfer coefficients \citep[i.e.,][]{Paxton2015, Rosales, Soydugan} were used in calculations as 0.4, 0.1, and 0.1 for $\alpha$, $\beta$, and $\gamma$, respectively.

\begin{figure*}[!t]
    \centering
    \includegraphics[width=.7\linewidth]{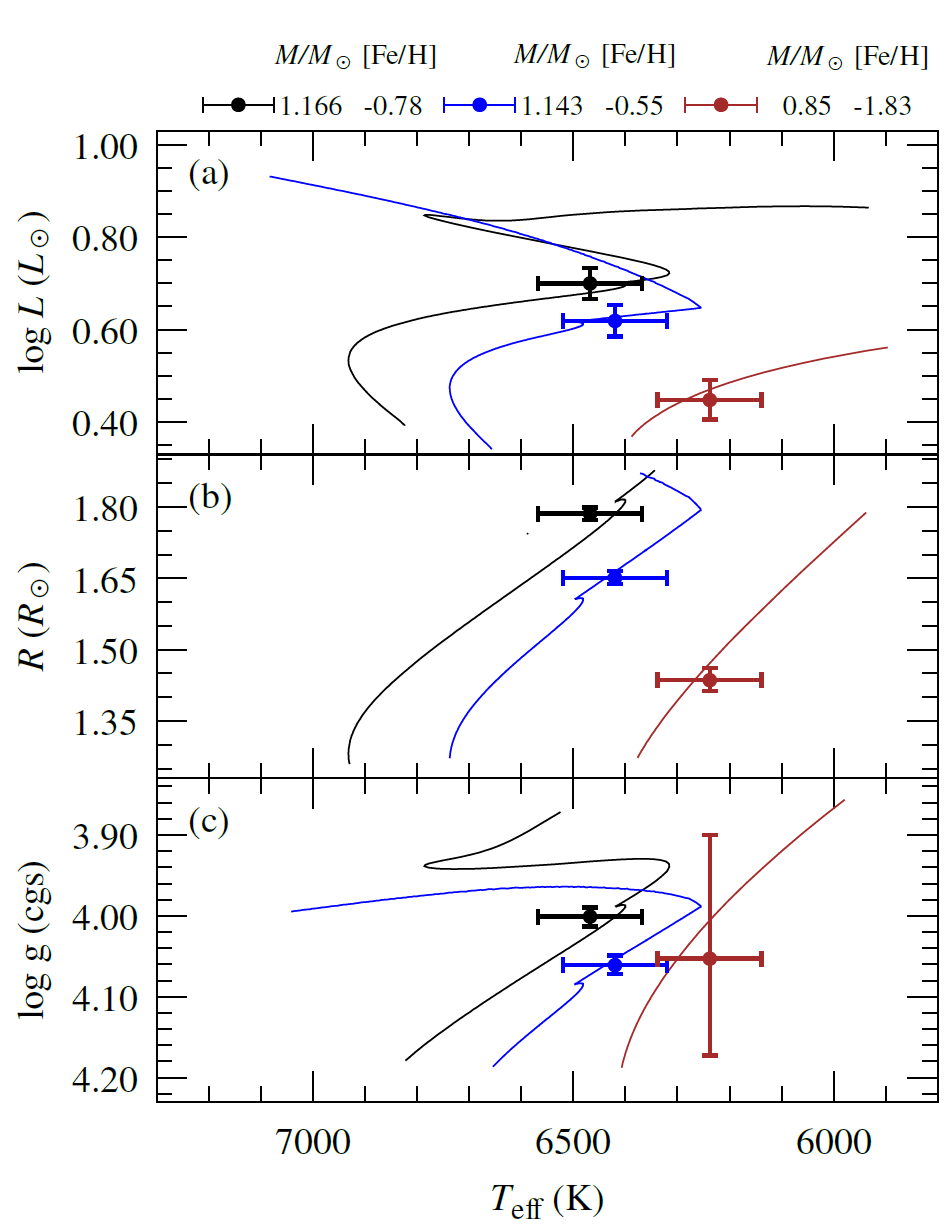}
    \caption{Positions of the primary (Aa), the secondary (Ab) and the tertiary (B) components of CN Lyn on evolutionary tracks in (a) $\log{L} \times T_{\rm eff}$, (b) $R \times T_{\rm eff}$, and (c) $\log g \times T_{\rm eff} $ planes, respectively. The different colored curves show the MESA evolutionary tracks of the derived metallicities (see Table ~\protect\ref{tab:parameters}) for the calculated masses. Red, blue, and brown colors represent the evolutionary track of each component: the primary, the secondary, and tertiary, respectively. The left turn in the blue tracks represents the Ab component starting to accrete mass from the Aa component.}
    \label{fig:all}
\end{figure*}

\begin{figure}[t!]
\centering
	\includegraphics[width=0.95\columnwidth]{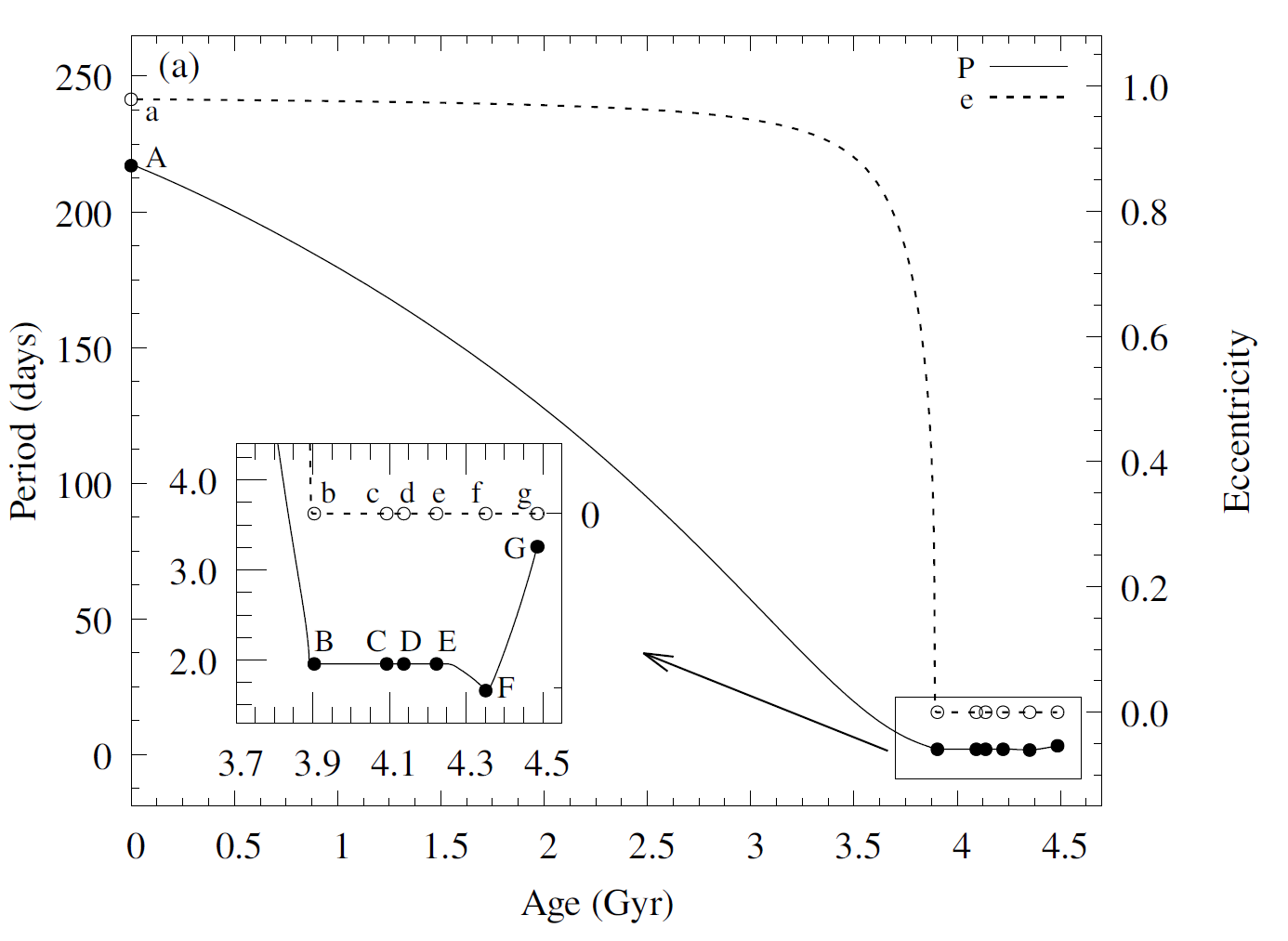}
 \includegraphics[width=0.95\columnwidth]{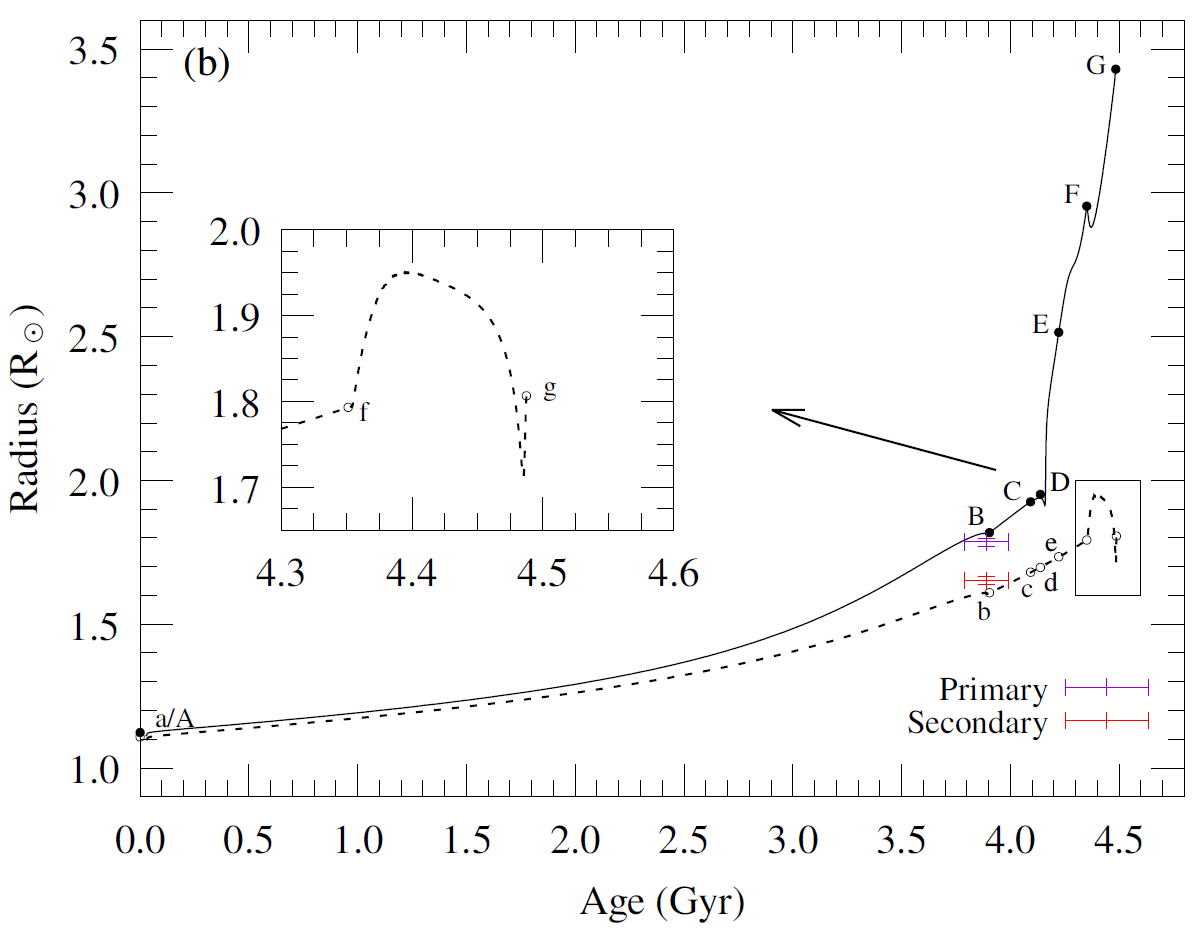}\vspace{-5pt}
     \caption{Change of orbital parameters (a), and radius of the components (b) of Aab components of CN Lyn with time.}
    \label{fig:radi}
\end{figure}

\begin{figure}[t!]
    \centering
    \includegraphics[width=\linewidth]{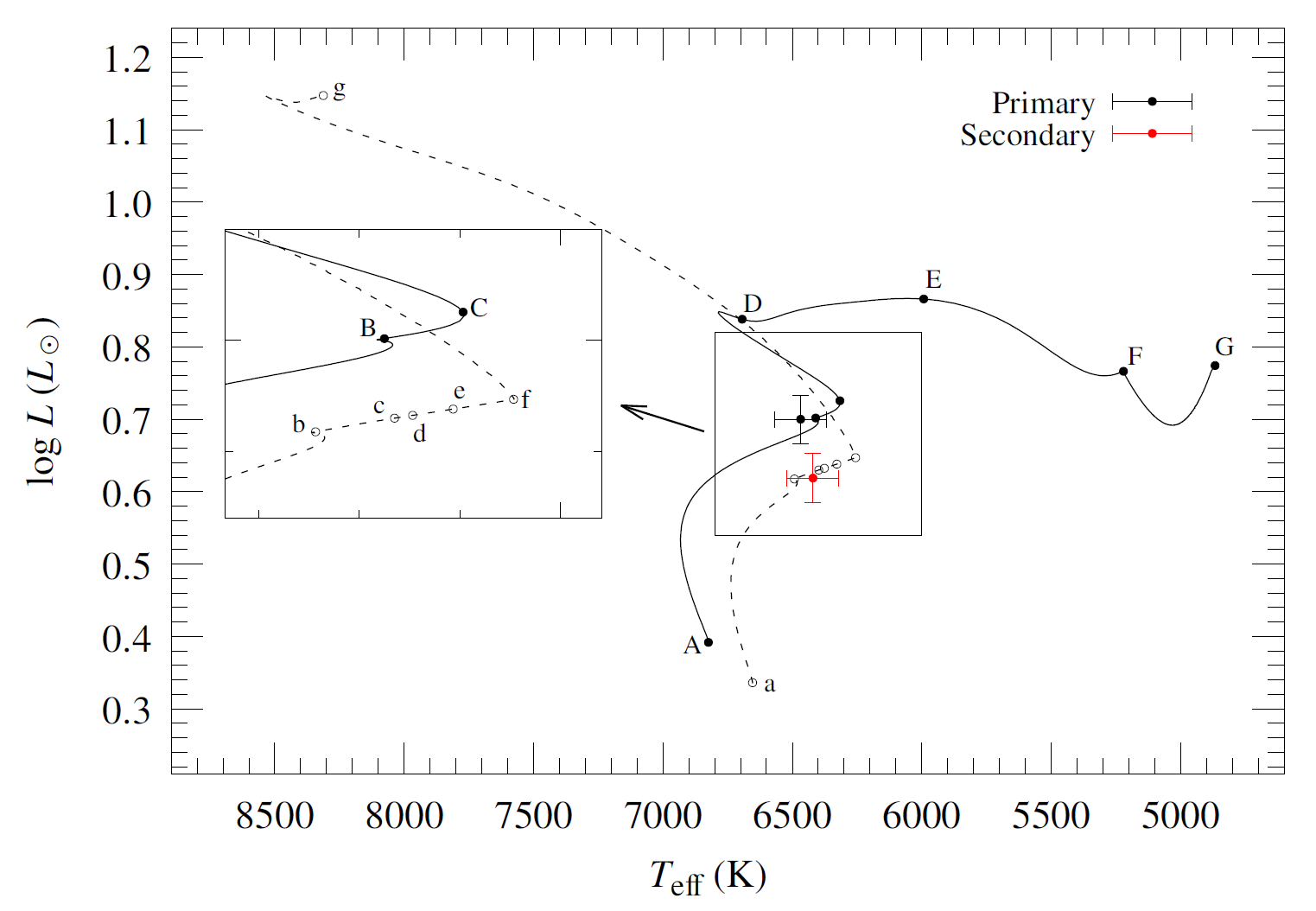}
    \caption{Evolutionary tracks of the primary (Aa) and the secondary (Ab) components of CN Lyn in $\log{L/L_\odot} \times T_{\rm eff}$ plane.}
    \label{fig:HR}
\end{figure}

Our analysis indicates that the age of the A system is $3.89_{-0.10}^{+0.10}$ Gyr and mass transfer in the system will start in about 460 Myr (see Table \ref{tab:CN_Lyn_Aab}). The current position of the components of Aab is given in Figure \ref{fig:all}. Changes in orbital parameters and radii of the components during the evolution are presented in Figure \ref{fig:radi}. Detailed evolution of both components with timetables is given in Table \ref{tab:CN_Lyn_Aab} and shown in Figure \ref{fig:HR}.


\begin{table*}[!ht]
\renewcommand{\arraystretch}{1}
\setlength{\tabcolsep}{3pt}
\centering
\scriptsize
    \caption{Detailed evolution of Aab with time-stamps.}
        \begin{tabular}{llccccccccc}
    \toprule
        Mark & \multirow{2}*{Evolutionary Status} & Age & $P$ & \multirow{2}*{$e$} & \multicolumn{3}{c}{Primary} & \multicolumn{3}{c}{Secondary} \\
       \cline{6-11}
         (Pri/Sec) & & (Myr) & (day) &  & $T_\mathrm{eff}$ (K) & $\log L$ ($L_\odot$)  & $R$ ($R_\odot$) & $T_\mathrm{eff}$ (K) & $\log L$ ($L_\odot$)  & $R$ ($R_\odot$)\\
        \hline
                     A/a     & ZAMS                              & 0     & 217.0     & 0.9781 & 6824     & 0.392 & 1.123 & 6653 & 0.336 & 1.108 \\
                     B/b     & Circularisation of orbit          & 3904  & 1.956     & 0      & 6410     & 0.702 & 1.819 & 6492 & 0.618 & 1.609 \\
                     C/c     & Primary core contraction          & 4093  & 1.956     & 0      & 6316     & 0.726 & 1.926 & 6398 & 0.630 & 1.681 \\
                     D/d     & Primary TAMS                      & 4138  & 1.956     & 0      & 6694     & 0.838 & 1.952 & 6376 & 0.632 & 1.697 \\
                     E/e     & Primary thin H shell burning      & 4223  & 1.956     & 0      & 5992     & 0.866 & 2.515 & 6328 & 0.638 & 1.735 \\
                     F/f     & Starting of mass transfer         & 4351  & 1.660     & 0      & 5220     & 0.766 & 2.954 & 6256 & 0.647 & 1.793 \\
                     G/g     & Secondary TAMS                    & 4485  & 3.259     & 0      & 4866     & 0.774 & 3.430 & 8312 & 1.147 & 1.807 \\
        \bottomrule
    \end{tabular} 
    \label{tab:CN_Lyn_Aab}
\end{table*}

\newpage 
\subsection{Evolution Analysis of B}

The chemical properties of component B are significantly different from those of Aa and Ab. This suggests that B did not form in the same molecular cloud as Aa and Ab but was instead captured by A at some point during its Galactic orbit. Thus, we approached evolution analysis for the B component as a single star. We used \texttt{star} module for calculating the evolutionary scenario of the B component. Our analysis points out that the B component is in the main-sequence band and very close to reaching terminal age main sequence (TAMS), and that its age is $12.5\pm 2.5$ Gyr. The evolutionary track and the position of the B component with other components, Aa and Ab, are shown in Figure \ref{fig:all}. 


\section{Kinematics and Galactic Orbit Parameters} \label{sec:kinematics}

Thanks to the precise astrometric measurements of the {\it Gaia} satellite, the kinematic and dynamical orbital parameters of relatively nearby stars can be calculated very accurately and precisely. In this study, the space velocity components and Galactic orbital parameters of CN Lyn were calculated using the trigonometric parallax ($\varpi$) and proper motion components ($\mu_{\alpha}\cos\delta, \mu_{\delta}$) provided from the {\it Gaia} DR3 catalogue \citep{Gaia_DR3} and the radial velocity ($V_{\gamma}$) data of the center-of-mass of the binary system. The astrometric and spectroscopic data used in the calculations are listed in Table~\ref{tab:kinematic}. 

We employed the \texttt{galpy} software package introduced by \citet{Bovy_2015} to compute the space velocity components of CN Lyn. The uncertainties related to these components were assessed utilizing the methodology outlined by \citet{Johnson_1987}. The space velocity components are affected by the position of the stars within the Galaxy and by natural deviations due to observations from the Sun. To reduce these biases, differential rotation corrections and local standard rest (LSR) adjustments were applied to each space velocity component. We corrected the differential rotational effects on CN Lyn using the equations detailed in \citet{Mihalas_1981}, resulting in velocity adjustments of $-0.17$ and $-0.48$ km s$^{-1}$ for the $U$ and $V$ components of the system, respectively. The $W$ component, being independent of differential rotation, did not require correction. For the LSR correction, we adopted the values from \citet{Coskunoglu_2011}, $(U, V, W)_{\odot}=(8.83\pm 0.24, 14.19\pm 0.34, 6.57\pm 0.21)$ km s$^{-1}$, and applied these to adjust the space velocity components after differential velocity corrections were made. The total space velocity ($S_{\rm LSR}$) of the system was computed using the relation $S_{\rm LSR}=\sqrt{U_{\rm LSR}^2+V_{\rm LSR}^2+W_{\rm LSR}^2}$, with results presented in Table \ref{tab:kinematic}. The kinematic method of \citet{Bensby2003} was taken into account in the tests of Galactic population classification according to the kinematic parameters of CN Lyn. The kinematic analyses gave the probability of CN Lyn belonging to the thin disk (D), thick disk (TD), and halo (H) populations as ($P_{\rm thin}, P_{\rm thick}, P_{\rm halo}) = (80.73, 19.20, 0.07)$\%. In other words, the probability ratio of the system being a member of the thick disk population to being a member of the thin disk population is $TD/D = 0.192$. This kinematic analysis indicates that it is highly likely that CN Lyn is a member of the thin disk.

\begin{table*}[t!]
    \setlength{\tabcolsep}{3pt}
    \renewcommand{\arraystretch}{1}
    \small
    \centering
    \caption{Astrometric measurements of CN Lyn and its radial velocity were used to calculate space velocity components and Galactic orbital parameters. The astrometric and spectroscopic data in the table are taken from the {\it Gaia} DR3 catalogue and this study, respectively.}
        \begin{tabular}{llccccccc}
        \hline
        \multicolumn{9}{c}{Input Parameters}\\
        \hline
        \hline

Star     &   $\alpha$ (J2000) & $\delta$ (J2000)  & $\mu_{\alpha}\cos\delta$ & 	$\mu_{\delta}$	 & \multicolumn{2}{c}{$\varpi$}       & \multicolumn{2}{c}{$V_{\rm \gamma}$}\\
         &  (hh:mm:ss)        & (dd:mm:ss)        &       (mas yr$^{-1}$)    &   (mas yr$^{-1}$) & \multicolumn{2}{c}{(mas)}          & \multicolumn{2}{c}{(km s$^{-1}$)}\\
\hline
CN Lyn   & 	08:01:37.20       & +38:44:58.41	  &  3.117$\pm$0.074         & 39.691$\pm$0.051   & \multicolumn{2}{c}{4.331$\pm$0.063}& \multicolumn{2}{c}{-14.90$\pm$0.10}\\	
\hline
\multicolumn{9}{c}{Output Parameters}\\
\hline
\hline
Star  &  $U_{\rm LSR}$     & $V_{\rm LSR}$  & $W_{\rm LSR}$  & $S_{\rm LSR}$	& $R_{\rm a}$	 & $R_{\rm p}$       & $Z_{\rm max}$ & $e$  \\
      &  (km s$^{-1}$)     & (km s$^{-1}$)  & (km s$^{-1}$)  & (km s$^{-1}$)    & (pc)           & (pc)              & (pc)          &      \\
\hline
CN Lyn   & 26.78$\pm$0.24  & 56.44$\pm$0.74	& 10.25$\pm$0.36 & 63.31$\pm$0.85   & 14\,813$\pm$107   & 8\,080$\pm$5        & 330$\pm$11    & 0.294$\pm$0.003\\
        \hline
        \end{tabular}
    \label{tab:kinematic}
\end{table*} 

\begin{figure}[ht!]
\centering\includegraphics[width=1\linewidth]{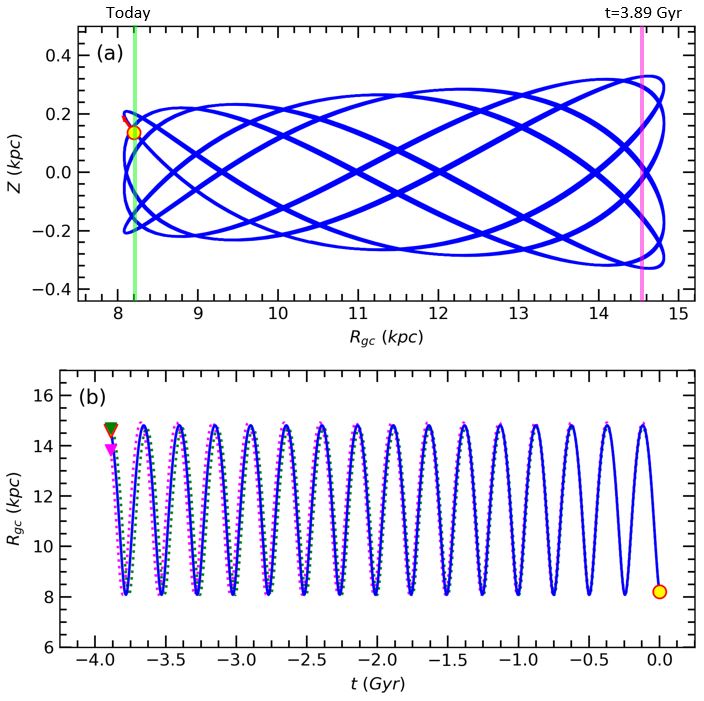}\vspace{-5pt}
\caption{The Galactic orbits and birth radii of CN Lyn in the $Z \times R_{\rm gc}$ (a) and $R_{\rm gc} \times t$ (b) diagrams. The filled yellow circles and triangles show the current and birth positions, respectively. The red arrow is the motion vector of CN Lyn today. The green and pink dotted lines show the orbit when errors in input parameters are considered, whereas the green and pink filled triangles represent the birth locations of the CN Lyn based on the lower and upper error estimates.}
\label{fig:galactic_orbits}
\end{figure}

We utilized the \texttt{galpy} code \citep{Bovy_2015} to determine the Galactic orbital parameters of CN Lyn. The Galactic potential required for these calculations was modeled using \texttt{MWPotential2014}, designed specifically for the Milky Way. To ensure closed orbits around the Galactic center, we simulated the system over a timescale of 3.9 Gyr in steps of 1 Myr. The Galactic orbital computations yielded several key parameters, including the apogalactic distance ($R_{\rm a}$), perigalactic distance ($R_{\rm p}$), maximum distance from the Galactic plane ($Z_{\rm max}$), and orbital eccentricity ($e$). The calculated Galactic orbital parameters are detailed in Table \ref{tab:kinematic}. The positions of the system relative to the Galactic center ($R_{\rm gc}$) and perpendicular to the Galactic plane ($Z$) at various time intervals are represented in Figure \ref{fig:galactic_orbits}a. The \texttt{galpy} analysis reveals that CN Lyn has a flattened Galactic orbit. Furthermore, the system's position $Z=114$ pc ($Z=d \times \sin b$) above the Galactic plane confirms that CN Lyn is probably part of the Milky Way's thin disk population \citep{Guctekin_2019}.

The Galactic orbits for the CN Lyn are shown in the $Z \times R_{\rm gc}$ and $R_{\rm gc} \times t$ diagrams in Figure \ref{fig:galactic_orbits}. These panels provide side views of CN Lyn's orbits, illustrating its distance from both the Galactic center and the Galactic plane \citep{Tasdemir_2023, Gokmen2023}. In Figure \ref{fig:galactic_orbits}b, the yellow-filled triangles and circles denote the birth and current positions of CN Lyn, respectively \citep{Yontan_2022, Akbaba2024}. The orbit of CN Lyn exhibits an eccentricity that does not exceed 0.30, with a maximum distance from the Galactic plane of $Z_{\rm max}=330\pm11$ pc. These findings suggest that CN Lyn is a member of the old thin disk of the Milky Way. The binary system's birthplace was conducted by tracing the binary system's age as $t=3.89$ Gyr backward using the \texttt{galpy} \citep{Yontan-Canbay_2023, Yucel2024}. This analysis determined the birth radius of the binary system to be $R_{\rm Birth}=14.59\pm0.86$ kpc which indicates that CN Lyn was formed in the metal-poor edge of the Milky Way's thin disk \citep{Gaia_DR3_metal, Katz2021}.

\section{Discussion} \label{sec:discussion}

This study's combination of information from spectroscopic, astrometric, and photometric data has led to the most detailed and accurate study of CN Lyn in the literature so far. A key technique was spectral disentanglement, allowing the chemical property of the three component stars to be estimated at high accuracy.  Our analysis indicates that the masses of each component (in the order Aa, Ab, and B) are $1.166_{-0.012}^{+0.013}$, $1.143_{-0.012}^{+0.013}$, and $0.85_{-0.23}^{+0.23}~M_{\odot}$. In the same order, the component radii are $1.786_{-0.014}^{+0.013}$, $1.651_{-0.013}^{+0.014}$, and $1.436_{-0.023}^{+0.026}~R_{\odot}$. The component temperatures are  \ppnumber{6411}$_{-100}^{+100}$, \ppnumber{6406}$_{-100}^{+100}$, and \ppnumber{6238}$_{-100}^{+100}$~K. Finally, the metallicities are $-0.78_{-0.02}^{+0.02}$, $-0.55_{-0.02}^{+0.03}$, and $-1.83_{-0.11}^{+0.09}$~dex. Our evolutionary analysis for the central binary (Aa, Ab) indicates that its initial orbit had a period of $\sim 217$ days and a very high eccentricity of nearly 0.98. The age of the A components is $3.89_{-0.10}^{+0.10}$ Gyr and is very close to the circularization time for the orbit. The primary component (Aa) will fill its Roche lobe in some 452 Myr and then start to transfer mass to the secondary component (Ab). Once the mass transfer has begun, it is expected that the secondary component (Ab) will reach the TAMS phase 135 Myr later. Investigating the evolution of systems containing close-mass components such as this one is important to better understand the small differences in the evolutionary process and their effects on physical states \citep[e.g.,][]{Southworth2013, Alicavus2022, Yucel2022, Alan2024}.

To set these results in context, we quickly review results from the literature. \cite{Marrese2004} calculated $1.04 \pm 0.02 M_\odot$ for both Aa and Ab masses, $R_{\mathrm{Aa}} = 1.80 \pm 0.21 R_\odot$, and $R_{\mathrm{Ab}} = 1.84 \pm 0.24 R_\odot$.  The current study was able to derive individual masses for the two stars which are both a little more massive (and outside the formal errors).  The radii from the current study are somewhat smaller than the estimates from \cite{Marrese2004}, but within the uncertainty ranges. These differences will be largely driven by the use of the more recent, and higher quality (\textit{TESS}) photometry, additional radial velocity data (ELODIE), and modeling the sinusoidal $\gamma$ velocity change caused by the third component (B). We therefore argue that the current results are a refinement of the earlier ones and in general agreement. \cite{Marrese2004} also estimated, via color analysis, the temperature of Aa as $T_\mathrm{eff} =$\ppnumber{6500}$\pm 250$ K and for Ab $T_\mathrm{eff} = $ \ppnumber{6455}$ \pm 260$ K. \cite{Liao2021} also estimated effective temperatures based on their spectroscopic observations. They obtained a low-resolution spectrum of CN Lyn at phase $\sim 0.5$ phase. Based on these data, \cite{Liao2021} concluded that $T_\mathrm{eff,2} =$ \ppnumber{6337}$\pm 37$ K, $\log{g} = 4.27 \pm 0.08$ cgs, and ${\rm [Fe/H]}_2 = -0.67 \pm 0.06$ dex. However, the analysis is contaminated with the spectrum of the third component (B). In contrast, the temperature analysis of the current study has been performed by, first, disentangling the spectrum of each component from composite spectra and then making spectroscopic analyses for each component. We contend that this has led to more accurate estimates.

\cite{Marrese2004} considered a $29 \pm 6\%$ contribution by component B to the modelled \textit{Hipparcos} photometry, stating that B's mass must be similar to those of the other two components. Based on this comment, \cite{Liao2021} calculated the orbital inclination angle of the third component to be $25^{\circ}$. In the present study, we first calculated the mass ratio of the third component (B) and the close binary system (A) via obtaining the spectroscopic orbit, leading to the estimation of orbital inclination of the B-component as $26^{\circ}\!\!.7_{-2.1}^{+2.1}$. \cite{Marrese2004} note the limited radial velocity data available to them for analysis. The current paper's orbital period is considerably shorter than that estimated by \cite{Liao2021}, while the estimated mass is also lower than estimates by previous studies. The increased data available to the current study allowed a more detailed analysis and refined estimates.

To determine whether the given triple star system (AabB) is hierarchical, we used the stability criterion for hierarchical triple systems. A hierarchical system is one in which the inner binary (Aa and Ab) is well-separated from the outer companion (B) that orbits the center of mass of the inner binary. According to the criterion outlined by \cite{Eggleton1995}, the ratio of the outer orbit’s semi-major axis (\(a_{\text{outer}}\)) to the inner orbit’s semi-major axis (\(a_{\text{inner}}\)) should be significantly large to ensure stability. Specifically, the ratio should satisfy the following relation:

\[ \frac{a_{\text{outer}}}{a_{\text{inner}}} > 5 \times \left( \frac{1 + e_{\text{outer}}}{1 - e_{\text{outer}}} \right)^{2/5} \times \left( \frac{m_{\text{outer}} + m_{\text{inner}}}{m_{\text{inner}}} \right)^{1/5} \]

\noindent where \(e_{\text{outer}}\) is the eccentricity of the outer orbit, \(m_{\text{outer}}\) is the mass of the outer component (B), and \(m_{\text{inner}}\) is the total mass of the inner binary (Aa + Ab). For our system, \(a_{\text{outer}} = 4.53\) AU, \(a_{\text{inner}} = 0.041\) AU, \(e_{\text{outer}} = 0.55\), \(m_{\text{inner}} = 1.166 + 1.143 = 2.309\) $M_\odot$, and \(m_{\text{outer}} = 0.85\) $M_\odot$. Using these values, we calculate the stability ratio and criterion. The ratio is \(\frac{a_{\text{outer}}}{a_{\text{inner}}} = \frac{2.86}{0.041} \approx 110.5\). The stability criterion is calculated as:

\[ 5 \times \left( \frac{1 + 0.55}{1 - 0.55} \right)^{2/5} \times \left( \frac{0.85 + 2.309}{2.309} \right)^{1/5} \approx 8.7 \]

\noindent Since \(110.5 > 8.7\), the system satisfies the stability criterion, indicating that it is a hierarchical triple system. This analysis demonstrates that the given system, with B orbiting the close binary Aa and Ab, meets the requirements for being considered hierarchical. Such systems are crucial for understanding the dynamic interactions and long-term stability of multiple star systems \citep{Eggleton1995, Tokovinin2018}.

\begin{figure*}[t!]
\centering\includegraphics[width=1\linewidth]{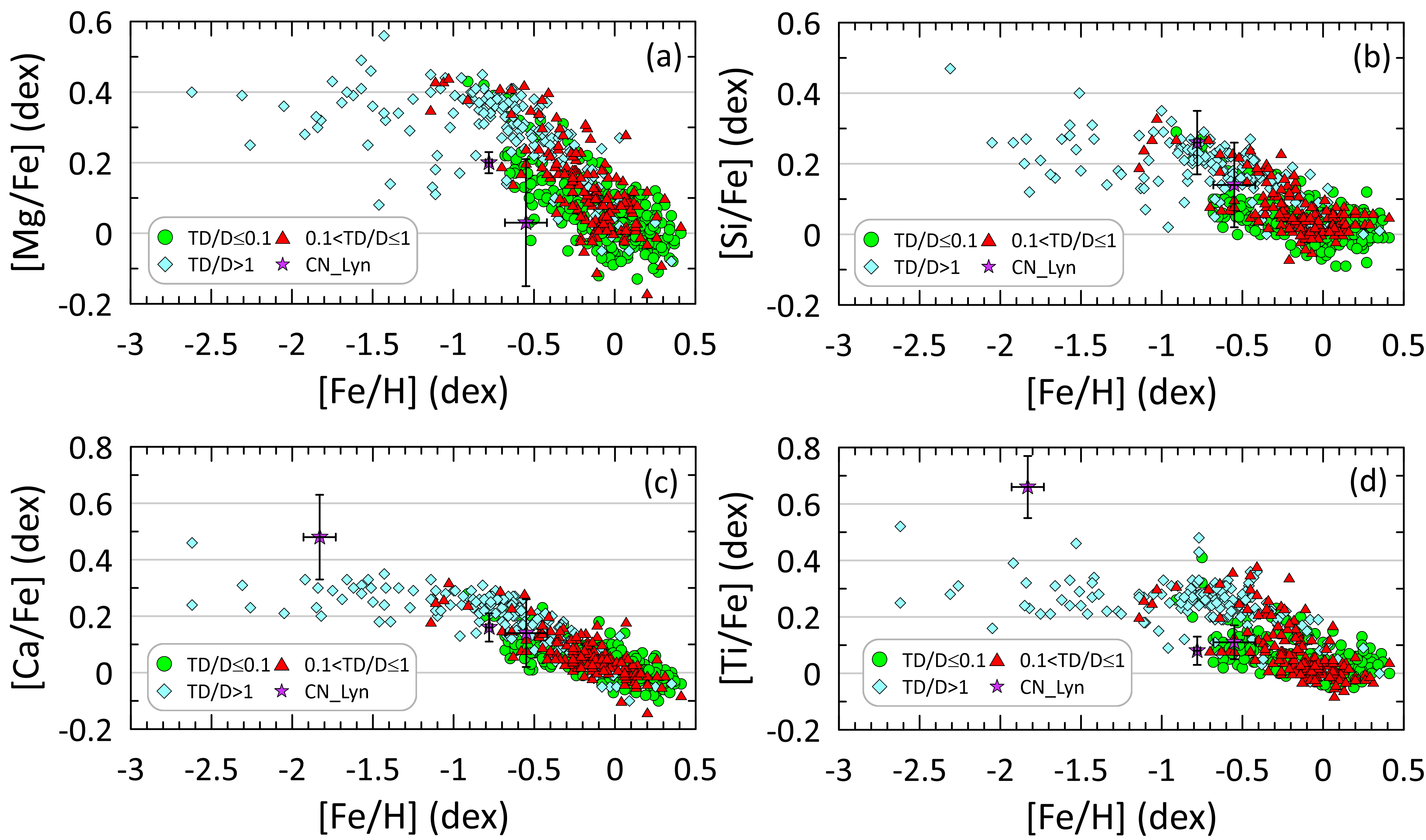}\vspace{-5pt}
\caption{Element abundances for the four $\alpha$ elements with respect to Fe. Panels show CN Lyn's component stars and \citet{Bensby2014}'s stars in the thin disk, thick disk, and halo populations classified according to kinematic criteria.}
\label{fig:Alpha_elements}
\end{figure*} 

Dynamical orbital analyses, taking into account the ages of the component stars in the central binary system (Aa, Ab) show that the system originated at the metal-poor edge of the Galactic disk, at a distance of $R_{\rm Birth}=14.59\pm 0.86$ kpc from the centre of the Galaxy. In the metal abundance analysis for the component stars, the mean metal abundance of the binary system was determined to be about $-0.65$ dex. This result is in agreement with the metal abundance of approximately $-0.45$ dex for 14.5 kpc given by \citet{Katz2021} in their ${\rm [Fe/H]} \times R$ diagram starting from the evolved stars in the APOGEE spectroscopic survey \citep{Majewski2017}. The metal abundance difference of up to 0.2 dex for the component stars in the binary system is also remarkable. This discrepancy may be due to the fact that the molecular cloud did not have a homogeneous metal abundance distribution during the formation of the binary system \citep[e.g.][]{Reipurth2007, Vogt2012, King2012}, atomic diffusion in stellar surface and interiors in the component stars \citep[e.g.][]{Yildiz2008, Dotter2017, Moedas2022, Liu2021}, or the metal-rich second component swallowed an exoplanet \citep{Liu2024, Saffe2024}.

Since the convective envelopes of Sun-like stars ($0.8<M_{\odot}<1.2$) are thin, a significant change in the metal abundance by atomic diffusion is expected for stars in the main-sequence band. The fact that the component stellar masses of CN Lyn in the A system are larger than $1 M_{\odot}$ and their mass ratio is close to unity ($q \sim 1$) makes it difficult to explain the differences in metal abundances by the diffusion mechanism \citep{Behmard2023}. However, the fact that the components in system A are very similar to the each other may provide important evidence that the metal abundance difference between the components ($\Delta$[Fe/H]=0.23 dex) was increased by one of the component stars swallowing an exoplanet. \citet{Oh2018} analyzed the stars HD 240429 (Kronos) and HD 240430 (Krios), determining that the metal abundance difference between the systems to be about 0.20 dex and showed that the metal-rich Kronos has a high probability of having `swallowed' an exoplanet of 15-50 $M_{\oplus}$. A possible explanation for the difference in metal abundances between the components in the system is that the component stars in the A system may occur in different chemodynamic field regions of the cloud in which they formed \citep{Brunt2009, Behmard2023}. A recent study showing that they are born in regions of the same cloud with different metal abundances was presented by \citet{Dursun2024}. Performing spectral energy distribution (SED) analyses of the stars in the open cluster NGC 188, \citet{Dursun2024} showed that the stars with different luminosities in the cluster may have $\Delta$[Fe/H]=0.27 dex. \citet{Dursun2024} also showed that the metal abundance variation within the cluster is 0.21 dex by comparing 12 common stars whose metal abundances were determined by spectral analyses by \citet{Jacobson2011}. The third component (B) has an orbital period $P=3\,130\pm 78$ days about the central pair of stars, mass $0.85\pm 0.23 M_{\odot}$, metal abundance $-1.81\pm 0.10$ dex, and $12.5\pm 2.5$ Gyr. Therefore, its chemical composition and age do not appear to be consistent with those of the other two component stars (A system).

In determining the Galactic population type of CN Lyn, the alpha and iron element abundances and kinematic data of the component stars were taken into account. Since there is no detailed elemental abundance analysis of CN Lyn in the literature, the elemental abundances calculated for the three component stars and the kinematic findings determined for the system were evaluated together with the 714 F and G spectral type stars selected by \citet{Bensby2014} from the solar neighborhood. In this study, the abundances of four alpha elements (Mg, Si, Ca, and Ti) identified in the spectra of three components in the CN Lyn are plotted on the [X/Fe]$\times$[Fe/H] chemical planes (Figure \ref{fig:Alpha_elements}) together with the abundances of \citet{Bensby2014}. The stars in the \citet{Bensby2014} sample are classified into three groups according to kinematic criteria as following high probability thin disk ($TD/D\leq 0.1$), low probability thin disk ($0.1<TD/D\leq 1$) and low/high probability thick disk or halo ($TD/D>1$) stars. While four alpha abundances were determined for the component stars in system A, only Ca and Ti abundances were detected in the chemical planes for component B. On the chemical planes, it was found that the stars in system A are located in the region where the thin disk members of \citet{Bensby2014} are located, while the metal-poor star denoted by component B is located in the position of thick disk/halo stars. In this case, the chemical abundances and kinematic data suggest that the stars in the A system belong to the old thin disk and component B to the halo population. Moreover, the fact that the component stars in system A are relatively poor in terms of alpha elements and iron abundances indicates that the molecular cloud forming the system was formed by Type Ia supernovae \citep[e.g.,][]{Gratton2000, Matteucci2009, Maoz2017}, while the rich Ca and Ti element abundances of component B suggest that it was formed by Type II supernovae \citep[e.g.,][]{Wyse1992, Matteucci2021}.

System A may have captured this third component in a region of weak gravitational interactions far beyond the Galactic centre. In conclusion, the fact that the tertiary star (B) of the CN Lyn is older and metal-poor compared to the close orbit two components (A) suggests that the tertiary component was captured by a gravitational effect sometime after the formation of the Aab-system. The absolute parameters calculated for the component stars in CN Lyn in this study will contribute to the understanding of the formation region and dynamical evolution of triple stars in the Galaxy. In addition, the results of this study will make a significant contribution to the study of the evolution of triple systems and Galactic archaeology. Further precise spectroscopic observations of the CN Lyn system would allow a more accurate determination of the center of mass of the system, the velocity variation, and the chemical abundances of the third component, thereby enabling a more reliable determination of the nature of the third body, and the understanding of the capture process.


\section*{Acknowledgments}
We thank the anonymous referee for their insightful and constructive suggestions, which significantly improved the paper. We thank T\"{U}B{\.{I}}TAK for funding this research under project number 123C161. This study was funded by Scientific Research Projects Coordination Unit of Istanbul University. Project number: MAB-2021-37903. We thank the following colleagues for their valuable discussions to make this study better in the following topics: Dr.\ Ulisse Munari for providing ASIAGO spectra, Prof.\ Petr Hadrava for disentangling spectra with KOREL, Dr.\ Kyle Conroy for PHOEBE calculations, Dr.\ Corrado Boeche for SP\_Ace usage, and lastly Dr.\ Olcay Plevne for TRUBA data preparations. The numerical calculations reported in this paper were partially performed at T\"{U}B{\.{I}}TAK ULAKB{\.{I}}M, the High Performance and Grid Computing Center (TRUBA resources). We made use of spectral data retrieved from the ELODIE archive at Observatoire de Haute-Provence (OHP). This research has made use of NASA's Astrophysics Data System. The VizieR and Simbad databases at CDS, Strasbourg, France were invaluable for the project as were data from the European Space Agency (ESA) mission \emph{Gaia}\footnote{https://www.cosmos.esa.int/gaia}, processed by the \emph{Gaia} Data Processing and Analysis Consortium (DPAC)\footnote{https://www.cosmos.esa.int/web/gaia/dpac/consortium}. Funding for DPAC has been provided by national institutions, in particular, the institutions participating in the \emph{Gaia} Multilateral Agreement. This paper includes data collected with the TESS mission, obtained from the MAST data archive at the Space Telescope Science Institute (STScI). Funding for the TESS mission is provided by the NASA Explorer Program. STScI is operated by the Association of Universities for Research in Astronomy, Inc., under NASA contract NAS 5–26555.

\software{\texttt{SP\_Ace} \citep{space1,space2}, \texttt{ATLAS9} \citep{Castelli2004}, \texttt{SPECTRUM} \citep{Gray1994}, \texttt{PHOEBE} \citep{phoebe1,phoebe2,phoebe3,phoebe4,phoebe5}, \texttt{Astropy} \citep{astro1,astro2,astro3}, \texttt{corner} \citep{corner}, \texttt{Matplotlib} \citep{matplotlib}, \texttt{NumPy} \citep{numpy}, \texttt{SciPy} \citep{scipy}, \texttt{galpy} \citep{Bovy_2015}, \texttt{MWPotential2014} \citep{Bovy_2015}.} 

\newpage

\bibliography{reference}
\bibliographystyle{aasjournal}

\appendix

\renewcommand{\thefigure}{A\arabic{figure}}
\setcounter{figure}{0}

\begin{figure}
    \centering
    \includegraphics[width=\linewidth]{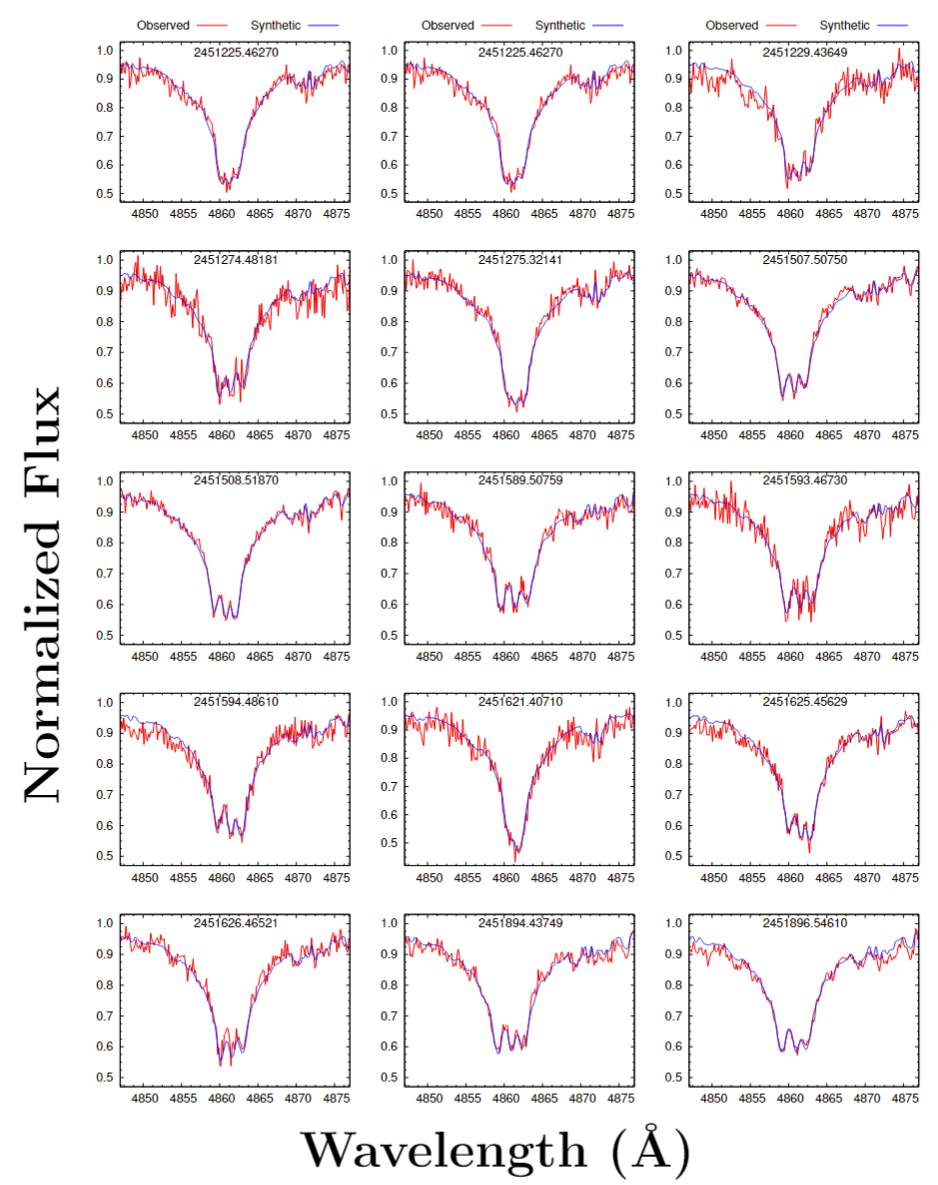}
    \label{fig:AA1}
\end{figure}
 \setcounter{figure}{0} 
\begin{figure}[t!]

    \centering
    \includegraphics[width=\linewidth]{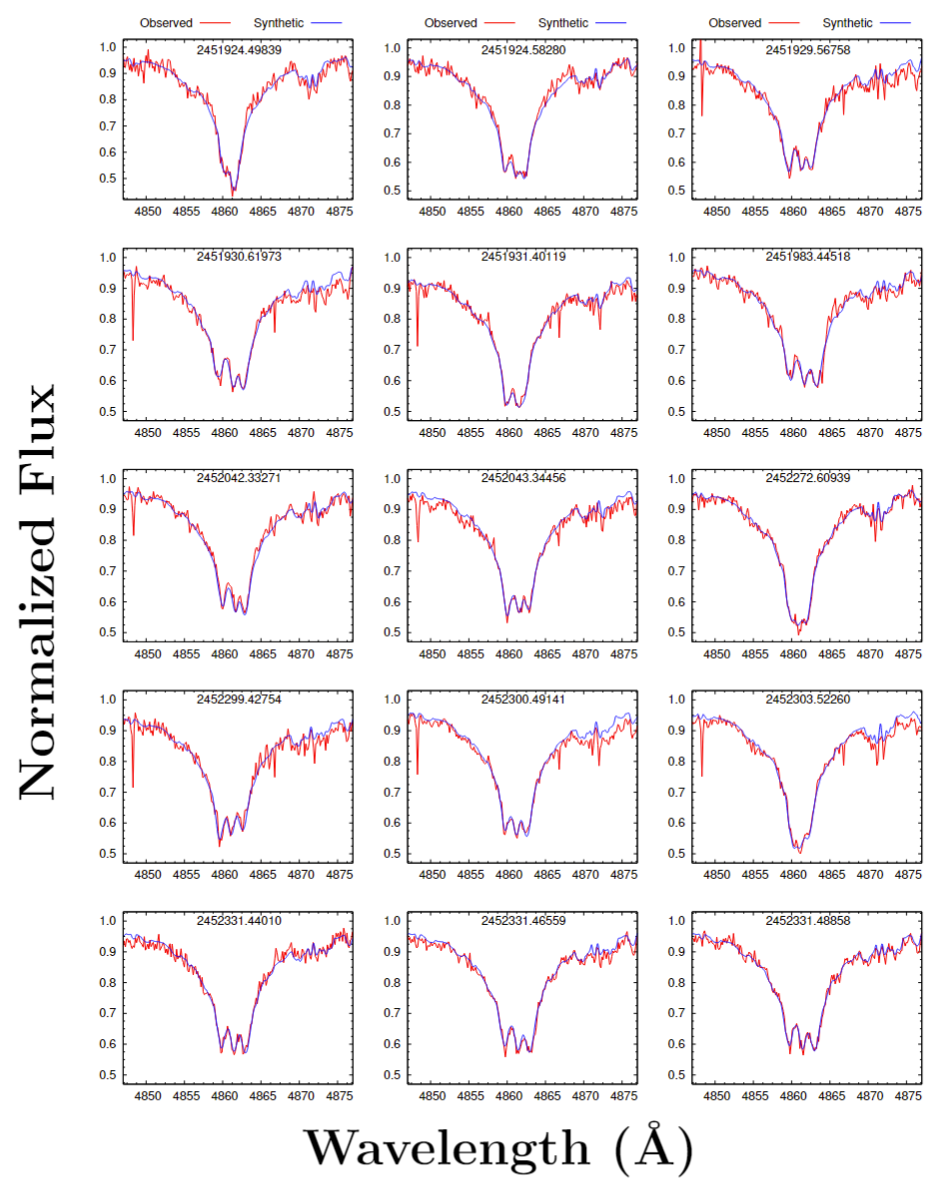}
\end{figure}

\begin{figure}[t!]
    \centering
    \includegraphics[width=\linewidth]{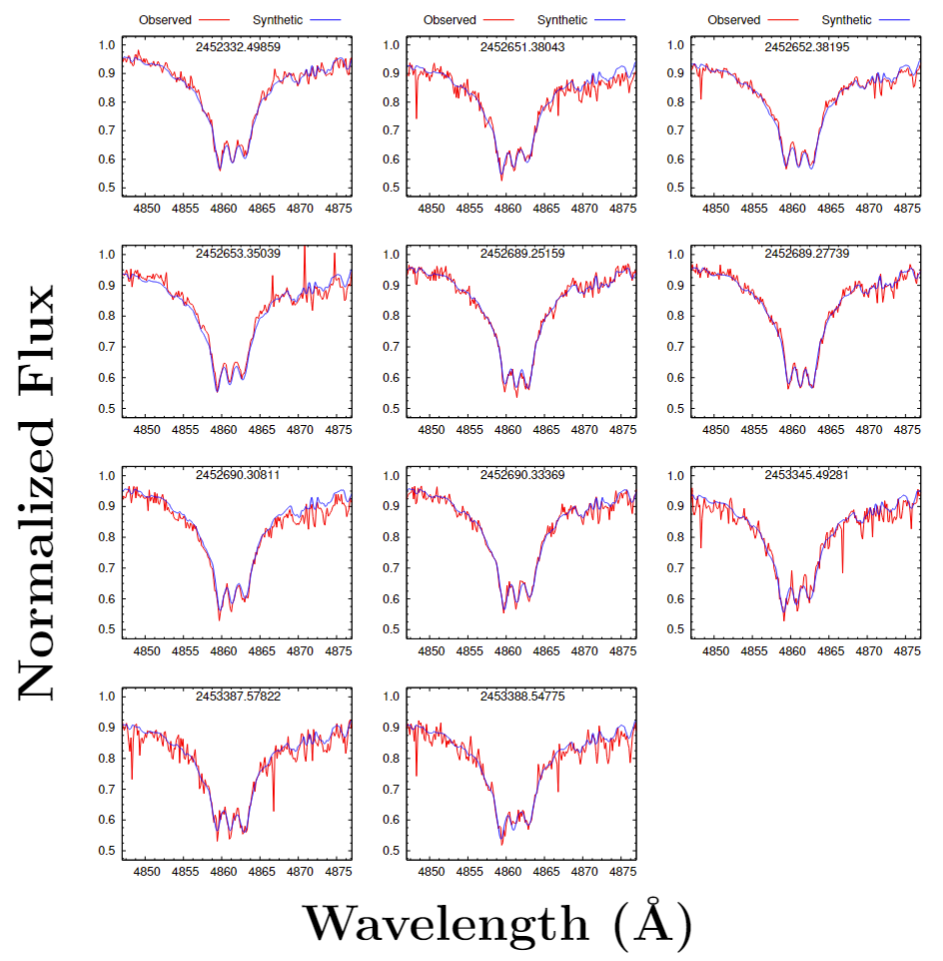}
      \caption{Models fits for RV measurements.}
\end{figure}

\end{document}